\documentstyle[epsfig,referee]{mn}

\def\prob{{\it Pr}}
\def\yr{{\rm yr}}
\def\pc{{\rm pc}}
\def\cm{{\rm cm}}

\title{The tidal disruption rate in dense galactic cusps containing a
  supermassive binary black hole}
\author[P. B. Ivanov, A.G. Polnarev, P. Saha]
       {P. B. Ivanov
\\ Astronomy Unit, School of Mathematical Sciences, Queen Mary,
University of London, UK \\
Astro Space Center of PN Lebedev Physical 
Institute, Moscow, Russia 
\newauthor A. G. Polnarev
\\ Astronomy Unit, School of Mathematical Sciences, Queen Mary,
University of London, UK
\newauthor P. Saha
\\ Astronomy Unit, School of Mathematical Sciences, Queen Mary,
University of London, UK
}

\pagerange{\pageref{firstpage}--\pageref{lastpage}}
\pubyear{1994}

\begin{document}

\maketitle

\label{firstpage}

\begin{abstract}

We consider the problem of tidal disruption of stars in the centre of
a galaxy containing a supermassive binary black hole with unequal
masses. We assume that over the separation distance between the black
holes the gravitational potential is dominated by the more massive
primary black hole.  Also, we assume that the number density of stars
is concentric with the primary black hole and has a power-law cusp. We
show that the bulk of stars with a small angular-momentum component
normal to the black-hole binary orbit can reach a small value of total
angular momentum through secular evolution in the gravitational field
of the binary, and hence they can be tidally disrupted by the larger
black hole.  This effect is analogous to the so-called Kozai effect
(Kozai 1962, Lidov 1961, 1962) well known in celestial mechanics. 
We develop an analytical theory for the secular evolution of 
the stellar orbits and calculate the rate of tidal disruption. 
We confront our analytical theory with a simple numerical model 
and find very good agreement.

Our results show that for primary black-hole mass 
$\sim 10^{6}-10^{7}M_\odot$,
the black-hole mass-ratio $q > 10^{-2}$, cusp size $\sim 1\pc$, the
tidal disruption rate can be as large as $\sim 10^{-2}-1M_\odot/\yr$. 
This is at least $10^{2}-10^4$ 
times larger than estimated for the case of a single
supermassive black hole. The duration of the phase of enhanced tidal
disruption is determined by the dynamical-friction time scale, and it
is rather short: $\sim 10^{5} \yr$.  The dependence of the tidal
disruption rate on the mass ratio, as well as on the size of the cusp,
is also discussed.

\end{abstract}


\section{Introduction}

The problem of the tidal disruption of the stars by a supermassive
black hole has attracted the attention of many researches starting
from the pioneering paper of Hills 1975, who proposed that the stellar
gas released from the tidally disrupted star could power quasars and
AGN's. Later, it has been realized that the tidal disruption event
could manifest itself as a luminosity flare in the centre of a galaxy
(e.g., Gurzadian \& Ozernoi 1981, Lacy et al.\ 1982, Rees 1988).  The
observational detection of these flares have recently been reported
(e.g., Komossa et al.~2003 and references therein).

In order to be tidally disrupted a star must have its orbital
pericentre sufficiently close to the central black hole, and
accordingly, its angular momentum must be sufficiently small. Thus,
the presence of the central black hole give rise to a `loss cone' in
the distribution of stars in angular-momentum space (e.g., Frank \&
Rees 1976), where the number of the stars is quickly exhausted due to
tidal disruption, and the process of refilling the loss cone
determines the rate of tidal disruption. Estimates of the rate of
tidal disruption are typically based on the assumption of two-body
gravitational interactions between the stars as the dominant process
of angular-momentum change. Since the relaxation time associated with
this process is large for realistic models of galactic centres, these
estimates give rather low values of the disruption rate, of the order
of $10^{-7}$ to $10^{-4}M_\odot/\yr$ depending on the properties of
the stellar population of the galactic centres and the mass of the
black hole residing in the centre (e.g., Frank \& Rees 1976, Lightman
\& Shapiro 1977, Shapiro \& Marchant 1978, Cohn \& Kulsrud 1978,
Magorrian \& Tremaine 1999, Syer \& Ulmer 1999, Wang \& Merritt 2004
and references therein).  It has been pointed out by Komberg 1968 and
by Polnarev and Rees 1994 that a secondary black hole of a smaller
mass could intrude in the galactic centre as a result of the process
of merging of two galaxies and perturb stellar trajectories, leading
to non-conservation of the angular momentum and an enhancement of the
tidal disruption rate (for a review of astrophysical processes occuring 
in galactic centres containing a supemassive black hole binary see e. g.
Merritt $\&$ Milosavljevic 2004 and references therein). 
The purpose of this paper is to calculate this
enhanced tidal disruption rate in a simple model of interaction
between the binary black hole and the stars.

We assume that the binary black hole consists of two black holes of
unequal masses moving around each other in circular orbits deep inside
the potential well of the primary black hole.  We take into account
the gravity of the central stellar cluster in the form of small
corrections to the motion of stars. It has been pointed out that when
the gravity is dominated by the central black hole, the number density
of stars increases toward the centre according to a power law and such
a distribution of the stars has been called `the cusp' (Peebles 1972,
Bachall \& Wolf 1976, Young 1980).  We use this power law distribution
as the initial condition for our problem. We consider only stars with
orbits within the orbit of the binary (which we will call the `inner
problem').  We show that in this case the main process leading to the
replenishment of the loss cone is a secular effect increasing the
eccentricities of the stars called `the Kozai effect' (Kozai 1962,
Lidov 1961, 1962).  The qualitative explanation of the process is very
simple.  Let us assume as a zeroth approximation that the stars move
on Keplerian orbits in the gravitational field of the primary black
hole and that gravitational interaction with the secondary black hole
introduces slow secular changes to the orbital elements.  Neglecting
the possibility of resonance interactions and considering the
evolution of the orbital parameters on a time scale scale much larger
than the orbital periods of the star and of the binary, it is
well-known that the secondary black hole can be approximated by its
orbital average, which is a mass ring.  Since the gravitational field
of a ring is static and axially symmetric, the energy and the
angular-momentum component normal to the ring plane are
conserved. However, in the course of the secular evolution the total
angular momentum varies with time, and it can be as small as the
constant normal component.  Since the number of stars whose normal
component of angular momentum lies within loss-cone size is much
larger than the number of stars whose total angular momentum lies
within the loss cone, the number of stars which can be tidally
disrupted is significantly increased by the presence of the secondary
black hole.

In the next Section we formulate the model problem, which takes into
account the secular evolution of the orbital elements of the stars
from both the secondary black hole and the stellar cluster. The
analytical solution of the secular problem is discussed in Section
3. The tidal disruption rate based on this solution is evaluated in
Section 4. In Section 5 we present a numerical approach to the problem
and check the validity of our analytical results. In Section 6 we use
the results obtained in the previous Sections to estimate the enhanced
tidal disruption rate in astrophysical systems.  This Section is
written in a fairly self-contained way, so readers interested in
astrophysical applications but not in the derivations of the relevant
equations can go directly to this Section. Section 7 presents our
conclusions and discussion.

\section{Formulation of the model problem}

As explained in the Introduction, we would like to know how many stars
are supplied on low angular-momentum orbits due to gravitational
interaction with the supermassive binary black hole consisting of a
primary black hole of mass $M$ residing in the centre of a galaxy and
an intruding secondary black hole with a lower mass $m$.  The
gravitational interaction of the binary black hole with stars of the
stellar cluster is an extremely complicated physical process which
involves dynamics of the stars in the field of two gravitating masses,
effects of a slow change of orbital elements of the binary black hole
due to dynamical friction and emission of gravitational radiation
(e.g., Polnarev \& Rees 1994, Quinlan \& Hernquist 1997, 
Merritt $\&$ Milosavljevic 2004 and references
therein) and due to interaction with an accretion disc (Ivanov,
Papaloizou \& Polnarev 1999), and effects caused by the gravitational
potential of the stellar cluster itself. Obviously, this problem
cannot be solved analytically. In order to make the problem
analytically tractable while retaining the main features expected in
an astrophysical situation, we consider a model where the orbital
elements of a star are evolving in Newtonian gravitational field of
two masses moving around each other on circular orbits. The separation
distance $D$ between two black holes remains constant with time. We
also take into account the most significant effect of the
gravitational field of the cluster on the stellar orbit, which is the
apsidal precession of the pericentre of the orbit. The rate of this
precession is calculated in Appendix A.  Apart from the apsidal
precession this model is the three-dimensional circular restricted
three-body problem of celestial mechanics.

Since we are interested in a statistical description of the stellar
orbits, we assume below that the stellar orbits are characterised by
phase space distribution function $f(\vec x, \vec v, t)$. Now let us
assume that at time $t=0$ the distribution function is not perturbed
by the secondary black hole and equals its unperturbed stationary
value $f_{0}(\vec x, \vec v)$. After this moment of time the
gravitational influence of the secondary hole is turned on and the
energies and angular momenta of the stars (defined in the frame
centred at the primary black hole) start to evolve due to
perturbations induced by the secondary.\footnote{We neglect the change
of the energies and angular momenta due to the usual gravitational
scattering of the stars.  This effect seems to be very slow comparing
to the effects determined by the secondary black hole.}  During the
evolution the specific angular momentum of some stars gets
smaller than a certain value $l_{T}$, corresponding to the tidal loss
cone.  These stars are assumed to be destroyed by the tidal forces of
the primary black hole and are not considered further.

Before going to details we would like to introduce some notation,
discuss some additional simplifying assumptions used later on, and
specify the initial conditions for the problem.

\subsection{Notation and additional simplifying assumptions}

When the semi-major axis of a star is sufficiently large, close encounters
of the stars with the secondary black hole are possible. These
encounters change the energy and the angular momentum of the stars in
a very complicated way, and the analytical treatment of these
encounters is rather difficult.  In this paper we do not consider
these complicated issues reserving them for future work and discuss
only the `inner problem' concentrating on the stars with semi-major
axis $a<D/2$.  In this case the orbit of the star does not cross the
orbit of the secondary black hole, and the theory of evolution of the
star's orbit is sufficiently simple (see below). In the following
discussion we introduce dimensionless variables associated with the
binary orbit: $\tilde a =a/D$, $\tilde L =L/\sqrt {GMD}$, etc. In
terms of these variables the problem depends only on two parameters:
the dimensionless absolute value of the angular momentum associated
with the tidal loss cone $l_{T}=L_{T}/\sqrt {GMD}$ and the mass ratio
$q=m/M$. Both parameters are assumed to be small, in the range
$q,l_{T} \sim 10^{-3}$ to $10^{-1}$ (see Section 6).

\subsection{The initial distribution function}
The number of stars supplied into the tidal loss cone obviously
depends on the form of the initial distribution function $f_{0}$.  In
this paper we are interested in effects occurring deep inside the
potential well of the primary black hole, where the number density of
stars grows steeply near the centre of the stellar cluster, forming
the so-called `cusp' in the stellar distribution. In order to model
this property of the central stellar clusters, we use the well known
power-law isotropic distribution function of the stars in the phase
space (e.g., Spitzer 1987, Bahcall and Wolf 1976, Lightman \& Shapiro
1977)
$$ f_{0}= CE^{p}, \eqno 1$$
where $C$ is a constant, $E=GM/r-v^{2}/2$ is the binding energy of a
star per unit mass in the field of the primary black hole of mass $M$,
and $v$ is the velocity of a star. According to theoretical estimates the parameter
$p < 3/2$ should be rather small, and
it can take two preferable values: $p=0$ (Peebles 1972, Young 1980) 
and $p=1/4$ (Bahcall and Wolf 1976). The value $p=0$ corresponds to adiabatic 
growth of the cusp in an unrelaxed stellar cluster containing a growing black hole
and the value $p=1/4$ corresponds to the situation when the characteristic relaxation
time is smaller than the life time of the system. Since the relaxation time is typically
rather large for the systems of interest we assume below $p=0$ for our estimates of the
tidal disruption rate in astrophysical systems (see Section 6). However we present results
from which the tidal disruption rate for the general case $p\ne 0$ can be easily obtained
(see Section 4). 

Also, let us note that after the binary black hole merging a significant
part of stars may be expelled from the system and the cusp may be destroyed. In fact, the
binary black hole scenario has been discussed to explaine rather shallow density profiles
observed in the centers of giant central galaxies situated in the centeres of galaxy clusters 
(e.g. Faber et al, 1997). We do not consider this more complicated situation in this Paper
assuming that the binary black hole is formed the first time in a 
particular galaxy of interest.

The number density of stars is obtained from (1) by
integration over velocity space:
$$n(r)= n_{a}\left({r\over r_{a}}\right)^{-(3/2+p)}, \eqno 2$$  
where $r_{a}$ is the radius of the cusp and $n_{a}$ is the number
density of stars at $r_{a}$. The constant $C$ in equation (1) is
obtained by integration of equation (1) over velocity space and
comparison the result of integration with equations (2). It is related
to $n_{a}$ as
$$C=2^{-5/2}\pi^{-1}\left({GM\over r_{a}}\right)^{-(p+3/2)}{n_{a}\over B(3/2;p+1)},
\eqno 3$$ 
where $B(x,y)$ is a Beta function
\footnote{Note the useful relations: $B(x,y)=B(y,x)={\Gamma (x) \Gamma(y)\over
\Gamma (x+y)}$, where $\Gamma (x) $ is the  Gamma function.}  
The total mass of stars inside a sphere of radius $r < r_{a}$ is 
given as
$$ M_{st}(r)=4\pi {m_{st} n_{a}\over (3/2-p)}r^{3}_{a}\left({r\over
r_{a}}\right)^{3/2-p}, \eqno 4$$ where $m_{st}$ is the star's mass and
we assume hereafter that the cusp consists of stars of equal masses
which are of the order of the solar mass $M_\odot$.  The total mass of
stars in the cusp must be of the order of the black hole mass $M$. In this
paper we assume that these two masses are equal:
$M_{st}(r_{a})=M$. This gives the normalisation condition:
$$n_{a}={(3/2-p)M\over 4m_{st} \pi  r_{a}^{3}}. \eqno 5$$
The initial distribution of the stars over the binding energies can be obtained from equation (1) by 
integration over the surface in the the phase space corresponding to a given value of 
the energy $E$ (e.g., Spitzer 1987):
$$N_{0}(E)=C_{E}E^{p-5/2}, \eqno 6$$
where $C_{E}=\sqrt{2}\left(\pi GM\right)^{3}C$. Below, instead of the distribution of the stars over the binding energy,
it will be
convenient to use the distribution over the semi-major axes $a$: 
$$N_{0}(a)=C_{a}a^{1/2-p}, \eqno 7$$
where $C_{a}=\pi^{3}2^{2-p}\left(GM\right)^{3/2+p}C$ and we use $E=GM/(2a)$.

We also need the distribution of stars in inclination angle $i$ and
angular momentum. To obtain this function let us introduce the
Cartesian coordinates with origin at the primary black hole and the
$z$ axis normal to the orbital plane of the binary black hole.  The
inclination angle $i$ is measured from the $z$ axis and the $z$
component of angular momentum, $L_{z}$, is related to its absolute as
$L_{z}=L \cos(i)$.  Assuming isotropy of the initial distribution
function $f_{0}$ and using equation (7) we immediately obtain the
initial distribution over the semi-major axis, inclinations and the
absolute values of angular momentum, $N_{0}(a,i,L)$
$$N_{0}(a,i,L)=N_{0}(a){L\sin (i)\over L_{a}^{2}}, \eqno 8$$ where
$L_{a}=\sqrt{GMa}$ is the angular momentum corresponding to the
circular orbit with radius $a$.  The initial distribution function
over semi-major axis, absolute values and the $z$-components of
angular momentum follows from equation (8):
$$N_{0}(a,L,L_{z})=N_{0}(a){1\over L_{a}^{2}}. \eqno 9$$

\section{The secular evolution}

The bulk of stars supplied to the loss cone will have had small
initial values of $L_z$ (see the next Section and Fig.~6 for numerical
results.)  This effect is determined by a particular feature of the
secular evolution of the star's orbital parameters in the
gravitational field of the binary.  For the Keplerian inner problem it
has been pointed out by Kozai 1962 and Lidov 1961, 1962 that although
the $a$ and $L_z$ are approximately conserved on time-scales much
larger than the orbital periods of the star and the binary, while $L$
varies with time. Therefore, a star with small $L_z$ initially can, in
the course of secular evolution, reach a small-enough value of $L$ to
be eventually tidally disrupted by the primary black hole.

Kozai and Lidov independently derived the evolution equations for the
orbital parameters using the methods of celestial mechanics. In
general, these equations have a rather complicated form. However, they
are simplified significantly when only the quadrupole contribution of
the gravitational potential of the secondary is taken into account
(e.g., Innanen, Zheng, Mikkola and Valtonen 1997, Kiseleva, Eggleton
\& Mikkola 1998). We use the evolution equations derived in these
papers and add an extra term describing the apsidal precession caused
by the gravity of the stellar cluster.

After doubly averaging over the periods of the star and the binary the
relevant equations take the form
$$T_{K}{de\over dt}=\frac52\epsilon e \sin^2 (i) \sin (2\omega), \eqno 10$$  
$$T_{K}{di\over dt}=-\frac54e^2 \sin (2i) \sin (2\omega)/\epsilon, \eqno 11$$  
$$T_{K}{d\omega \over dt}=(2(1-e^{2})+5\sin^2 
(\omega)(e^{2}-\sin^{2} (i)))/\epsilon-\kappa \epsilon, \eqno 12$$
$$T_{K}{d\chi \over dt}=-(1+e^{2}(5\sin^{2} (\omega )-1))\cos (i)/\epsilon. \eqno 13$$
Here $e$, $i$, $\omega$ and $\chi$ are the eccentricity, inclination,
the argument of pericentre and the longitude of ascending node,
respectively, and $\epsilon=\sqrt{1-e^2}$. $T_{K}$ gives the
characteristic time-scale of the Kozai effect:
$$T_{K}={4\over 3q}{\tilde a}^{-3/2}\sqrt{{D^{3}\over GM}}. \eqno 14$$

The last term in equation (12) describes the apsidal precession
induced by the stellar cluster (see Appendix A).  The parameter
$\kappa $ is given by
$$\kappa= K\left({T_{K}\over P_{orb}}\right) \left({M_{st}(a)\over M}\right)=
{2\over3\pi}K\left({M_{st}(a)\over m}\right), \eqno 15$$
where $P_{orb}=2\pi\sqrt{a^{3}/(GM)}$ is the orbital period of the star, 
$M_{st}(a)$ is the total mass of the stars inside a sphere of
the radius $a$, and the positive coefficient $K$ is calculated in
Appendix A:
$$K=2^{3/2-p} \sqrt \pi{\Gamma(1-p)\over \Gamma(3/2-p)}. \eqno 16$$ In
our estimates below we use $K=4\sqrt{2}$ corresponding to a cusp with
the parameter $p=0$ (see equations (1-9)). It is very important to
note that the apsidal precession caused by the stellar cluster has
negative sign while the sign of the apsidal precession caused by the
Kozai effect is positive. This stems from the fact that the
gravitational force acting from the cluster is always directed inward,
to the centre of the cluster, while the averaged gravitational force
acting from the secondary black hole in the inner problem is directed
outward.  The opposite directions of the gravitational forces result
in the opposite signs of corresponding precession rates.

The set of equations (10-13) has two integrals of motion: 
$$\Lambda=\epsilon \cos (i), \quad Q=e^{2}(5\sin^{2}(i)\sin^{2}(\omega )
+\kappa
-2). \eqno 17$$ It is obvious that the integral $\Lambda $ is
proportional to $L_z$.  Using the expressions (17) we can obtain the
evolutionary tracks of the dynamical system (10-13) on the plane $(e,
\omega )$:
$$\sin^{2}(\omega )={\epsilon^{2}(Q+(2-\kappa)e^{2})\over 5e^{2}
(\epsilon^{2} -\Lambda^{2})}. \eqno 18$$
Also, we can express the inclination $i$ and the argument of pericentre $\omega$ in 
terms of the eccentricity $e$ with help of the integrals (17), 
substitute the result in equation (12) and obtain a single evolution
equation for the eccentricity $e$: 
$$T_{K}e{de\over dt}=
\pm \sqrt{\lbrace ((3+\kappa +Q-5\Lambda^{2})
e^{2}-(3+\kappa)e^{4}-Q)(Q+(2-\kappa)e^{2})\rbrace}. \eqno 19$$
The solution of equation (19) can be expressed in terms of elliptical integrals (see below).

It can be shown that in the course of evolution of the dynamical
system (10-13) the change of the eccentricity with time is cyclic (we
will call these `Kozai cycles') and the argument of pericentre
$\omega$ can either evolve monotonically or librate around the values
$\omega=\pm \pi/2$. Accordingly, solutions of equations (10-13) can be
divided on two separate types of rotating $(r)$ and librating $(l)$
solutions. Let us discuss each types in turn.  As follows from
equation (10) the eccentricity $e$ as a function of time can have its
extremum only when $\omega=0$ or $\pi/2$. In the case of the solutions
of $(r)$ type both values of $\omega=0$ and $\omega=\pi/2$ are
possible. Let us call the values of eccentricity and inclination
corresponding to $\omega=0$ ($\omega=\pi/2$) as $e_{0}$ and $i_{0}$
($e_{1}$ and $i_{1}$).  The integrals (17) can be expressed either in
terms of $i_{0}$ and $e_{0}$ or $i_{1}$ and $e_{1}$. The corresponding
expressions can be equated to each other and we can obtain the
inclinations $i_{0}$ and $i_{1}$ as functions of $e_{0}$ and $e_{1}$
from the resulting equations:
$$\sin^{2}(i_{0})=
{1\over 5e_{1}}{(e_{1}^{2}-e_{0}^{2})\over (1-e_{0}^{2})}(2-\kappa+(3+\kappa)e_{1}^{2}),
\quad \sin^{2}(i_{1})={(2-\kappa)(e_{1}^{2}-e_{0}^{2})\over 5e_{1}^{2}}. \eqno 20$$
The integrals $\Lambda $ and $Q$ can also be expressed in terms of $e_{0}$ and $e_{1}$:
$$\Lambda^{2}={(1-e_{1}^{2})\over
5e_{1}^{2}}((3+\kappa)e_{1}^{2}+(2-\kappa)e_{0}^{2}), \quad
Q=-(2-\kappa)e_{0}^{2}. \eqno 21$$
As follows from equation (20) when $\kappa < \kappa_{\rm crit}=2$ the
maximal (minimal) value of the eccentricity is attained at $e=e_{1}$
($e=e_{0}$) and in the opposite case of a large $\kappa >
\kappa_{\rm crit}$ the maximal (minimal) value of eccentricity is equal to
$e_{0}$ ($e_{1}$).  In the last case the minimal value of the
eccentricity $e_{1}$ is always smaller than the critical value
$$e_{\rm crit}=\sqrt{\kappa -2\over \kappa +3},$$
and the maximal value $e_{0}$ is limited from above: $e_{0} < \min
\left(1, \sqrt{\kappa +3\over \kappa -2}e_{1}\right)$. It turns out
that when $\kappa > \kappa_{\rm crit}$ only a very small number of the
solutions of the $(r)$ type would lead to a significant increase of
the eccentricity, and accordingly, to a significant decrease of the
total angular momentum, during the secular evolution. Therefore, these
solutions are not relevant to our problem and will not be considered
further (see also the end of this Section).
   
When $\kappa < \kappa_{\rm crit}$ the solutions of the $(r)$ type
describe periodic changes of eccentricity and inclination, and the
minimal (maximal) value of the eccentricity corresponds to the maximal
(minimal) value of the inclination (see Fig.~1).

\begin{figure}
\vspace{8cm}\includegraphics{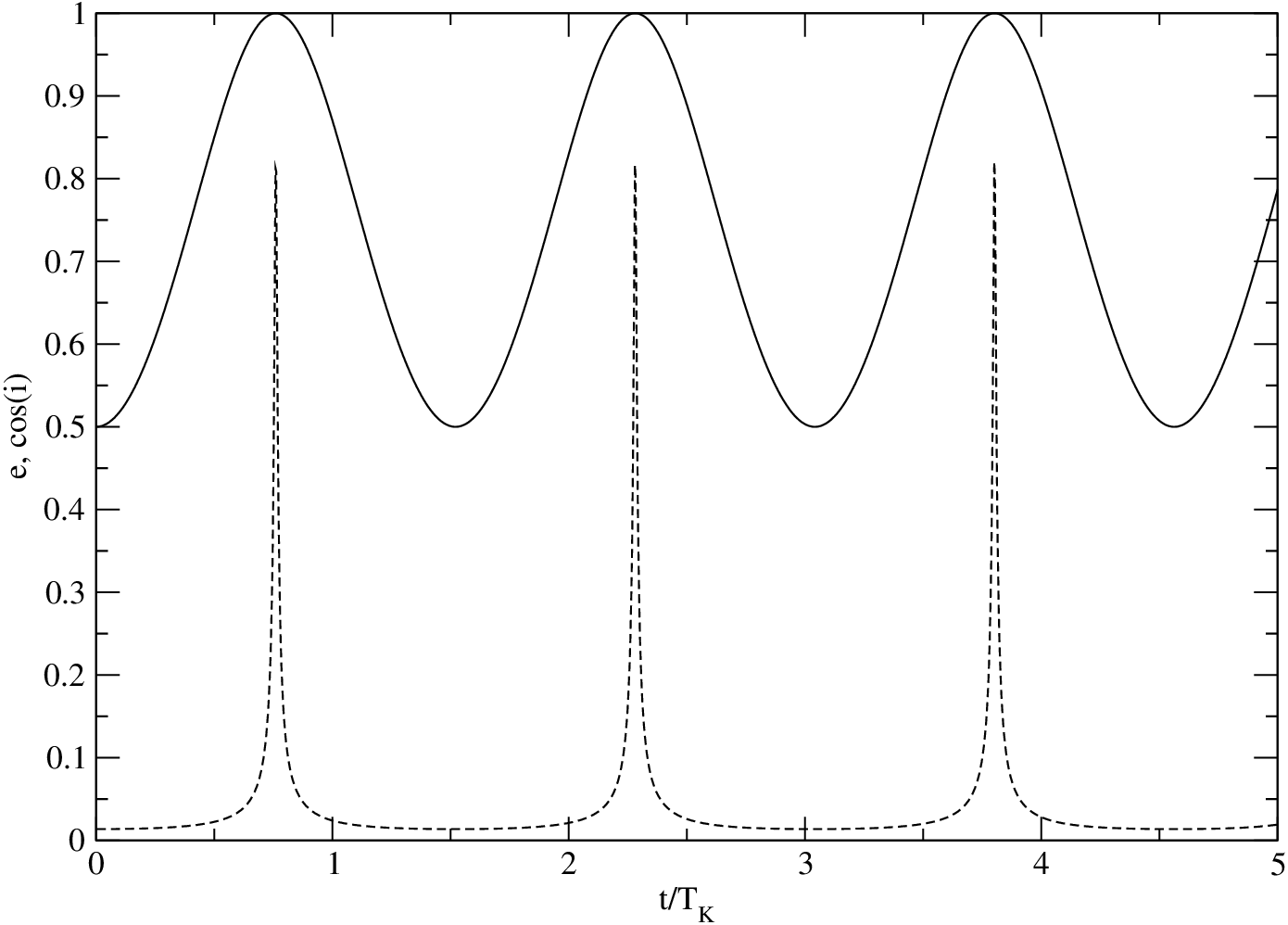}
\caption{An example of Kozai cycles. The dependencies of eccentricity
$e$ (solid curve) and $\cos(i)$ (dashed curve) on time (in units
$T_{K}$). The minimal (maximal) eccentricity is equal to $0.5$
($0.9999$).  The minimal value of $\cos(i) \approx 0$ ($ i \approx
\pi/2$) corresponds to the minimal value of the eccentricity, and the
maximal value of $\cos(i) \approx 0.8$ corresponds to the maximal
eccentricity. The period of the cycles $P_{K}\approx 1.5T_{K}$ is in
agreement with its theoretical value (see equation (24)). The case
shown corresponds to the Keplerian regime ($\kappa =0$) and the
evolution of $(r)$ type.}
\end{figure}

Substituting expressions (21) in equation (19) and integrating the
result we find the period of the Kozai cycles:
$$P_{K}={2\over\sqrt{(2-\kappa)((3+\kappa)e_{1}^{2}+(2-\kappa)
e_{0}^{2}/e_{1}^{2})}}{\bf K}(y)T_{K}, \eqno 22$$
where ${\bf K}(k)$ is the elliptic integral of first kind and
$$y=e_{1}^{2}\sqrt {{(3+\kappa)(1-e_{0}^{2}/e_{1}^{2})\over
(2-\kappa)e_{0}^{2}+(3+\kappa)e_{1}^{4}}}. \eqno 23$$
When $\kappa=0$ and $e_{0} \ll e_{1}$, $e_{1} \approx 1$ we have
$$P^{0}_{K}\approx \sqrt{2\over 3}\ln
\left(\sqrt {3\over 5}{4\over e_{0}}\right)T_{K}, \eqno 24$$
where we use the asymptotic relation ${\bf K}(k\rightarrow 1)\approx
\ln{{4\over k^{'}}}$, and $k^{'}=\sqrt{1-k^{2}}$. It is interesting to
note that when $e_{0} \approx e_{1} \approx 1$ the period $P^{0}_{k}$
is finite: $P^{0}_{K}\approx {\pi T_{K}\over \sqrt 10}\approx
T_{K}$. From equation (22) it also follows that when $\kappa
\rightarrow 2$ the period increases as $\propto 1/\sqrt{(2-\kappa)}$.

Now let us discuss the librating solutions where both minima and
maxima of the eccentricity are attained at $\omega=\pm \pi/2$. In this
case the corresponding inclination angles and the integrals (17) can
also be expressed in terms of the minimal and maximal eccentricities:
$$\sin^{2}(i_{\pm})={1\over 5}(2+3e^{2}_{\mp}-\kappa (1-e^{2}_{\mp})), \eqno 25$$
and
$$\Lambda^{2}={(3+\kappa)\over 5}(1-e_{-}^{2})(1-e_{+}^{2}), \quad
Q=(3+\kappa)e_{-}^{2}e_{+}^{2}, \eqno 26$$
where, by definition, $e_{-}$ ($e_{+}$) is the minimal (maximal) value
of the eccentricity and $i_{\pm}$ are the inclination angles
corresponding to $e_{\pm}$.  From equation (25) it follows that when
$\kappa > \kappa_{\rm crit}=2$ only sufficiently large eccentricities
are allowed: $e_{\pm} > e_{\rm crit}=\sqrt{\kappa -2\over \kappa +3}$.

Similar to the previous case the expression for the period of the
Kozai cycles is obtained by substitution of equation (26) in equation
(19) and integration of the resulting differential equation.  When
$\kappa < \kappa_{\rm crit}$ we have:
$$P_{K}={2\over e_{+}\sqrt{(3+\kappa)((3+\kappa)
e_{-}^{2}+2-\kappa)}}{\bf K}(y)T_{K}, \eqno 27$$
where
$$y=\sqrt{{(2-\kappa)(1-e_{-}^{2}/e_{+}^{2})\over
(3+\kappa)e_{-}^{2}+2-\kappa}}. \eqno 28$$
When $\kappa=0$, $e_{+}\sim 1$ and $e_{-} \ll 1$ we can obtain an
approximate expression for $P^{0}_{K}$ very similar to the expression
(24):
$$P_{K}^{0}\approx \sqrt{2\over 3}
\ln \left(\sqrt {3\over 5}{4\over e_{-}}\right)T_{K}, \eqno 29$$

In the opposite case $\kappa > \kappa_{\rm crit}$ we have
$$P_{K}={2\over (3+\kappa)e_{-}
\sqrt{e_{+}^{2}-e_{\rm crit}^{2}}}{\bf K}(y)T_{K}, \eqno 30$$
where
$$y={e_{\rm crit}\over e_{-}}\sqrt{{1-e_{-}^{2}/e_{+}^{2}
\over 1-e_{\rm crit}^{2}/e_{+}^{2}}}. \eqno 31$$
Note that when $\kappa=\kappa_{\rm crit}$ both expressions (27) and
(29) give the same result:
$$P_{K}={\pi \over 5e_{-}e_{+}}T_{K}. $$

When $e_{-} \rightarrow e_{\rm crit}$ the period diverges
logarithmically.  Assuming that $\delta =(e_{-}-e_{\rm crit})/e_{\rm
crit} \ll 1$ we find from equation (30)
$$P_{K}\approx  {2\over \sqrt{5(\kappa-2)}}
\ln \left({2\sqrt{{10\over (3+\kappa)\delta}}}\right) T_{K},  \eqno 32$$
where we also assume that $e_{+} \approx 1$ (see also equation
34). Note that the expression (32) is approximately valid only when
$\delta < \delta_{*}={5\over (\kappa +3)}$. Also, let us point out
that when both $e_{-}$ and $e_{+}$ are close to unity we can obtain
another approximate expression: $P_{K}\approx {\pi T_{K}\over
\sqrt{5(\kappa+3)}}$.
  
\begin{figure}
\vspace{8cm}\includegraphics{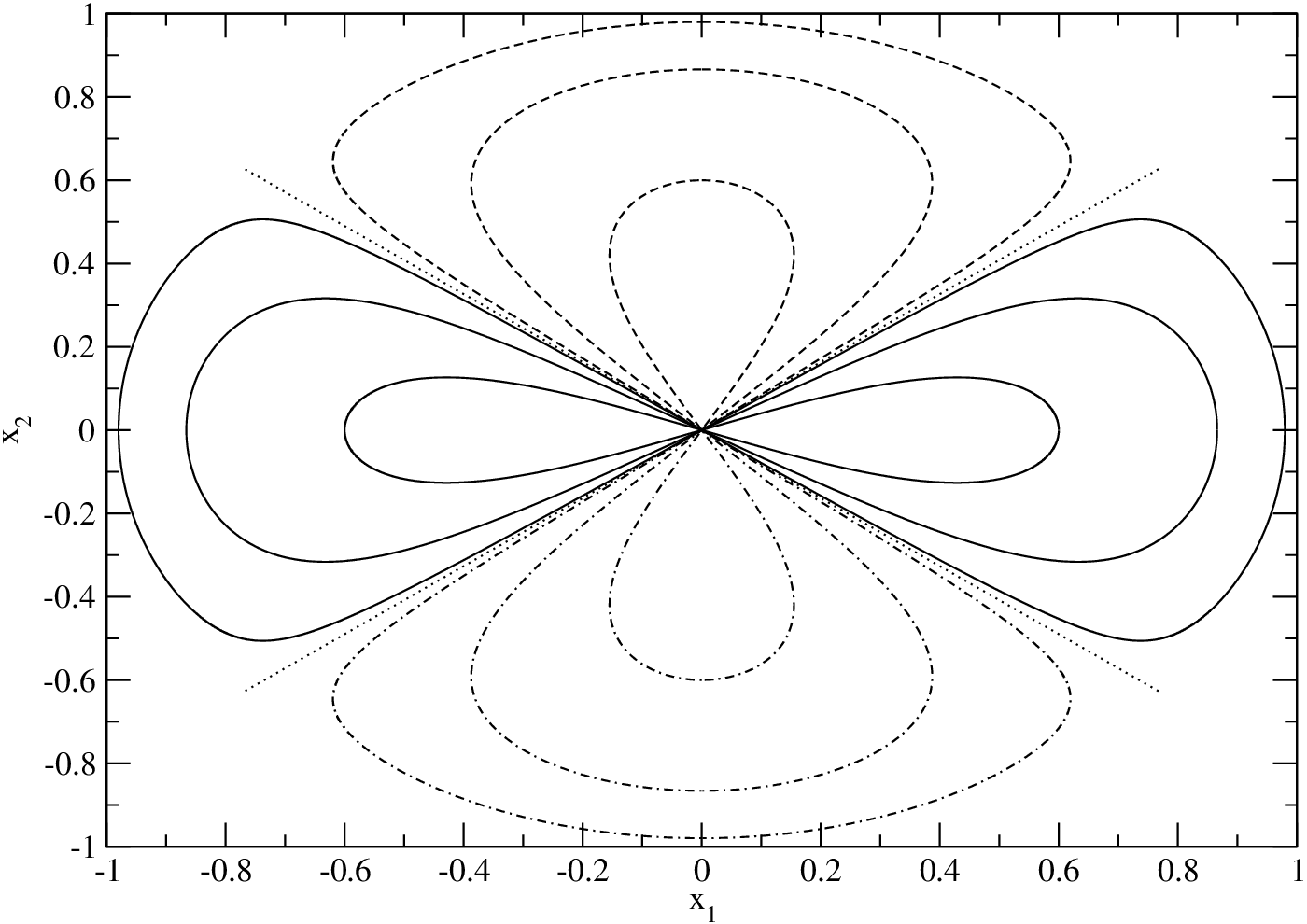}
\caption{We show the evolutionary tracks of the dynamical system
(10-12) on the plane $(x_{1},x_{2})$ for the purely Keplerian case
($\kappa=0$).  The value of the integral $\Lambda $ is taken to be
$1\cdot 10^{-3}$ for all curves. The solid curves represent the
solutions of $(r)$ type and the dashed and dot-dashed curves represent
the solutions of $(l)$ type. Different curves of the same type are
characterised by different possible minimal values of the
eccentricity, and respectively, by different maximal values of the
`radius' $\epsilon$. For the curves of $(r)$ type the minimal value of
the eccentricity $e_{0}$ is attained at $\omega=0$ and $\omega=\pi$
(when the curves cross the line $x_{2}=0$), and we present the cases
$e_{0}=0.2$, $0.5$ and $0.8$. The curves of $(l)$ type have the
minimal eccentricity $e_{-}$ at $\omega=\pm \pi/2$ (when the curves
cross the line $x_{1}=0$) and again we show the cases $e_{-}=0.2$,
$0.5$ and $0.8$. Two dotted lines separate the regions where the
librating and rotating regimes of evolution are possible (see the
text).}
\end{figure}

\begin{figure}
\vspace{8cm}\includegraphics{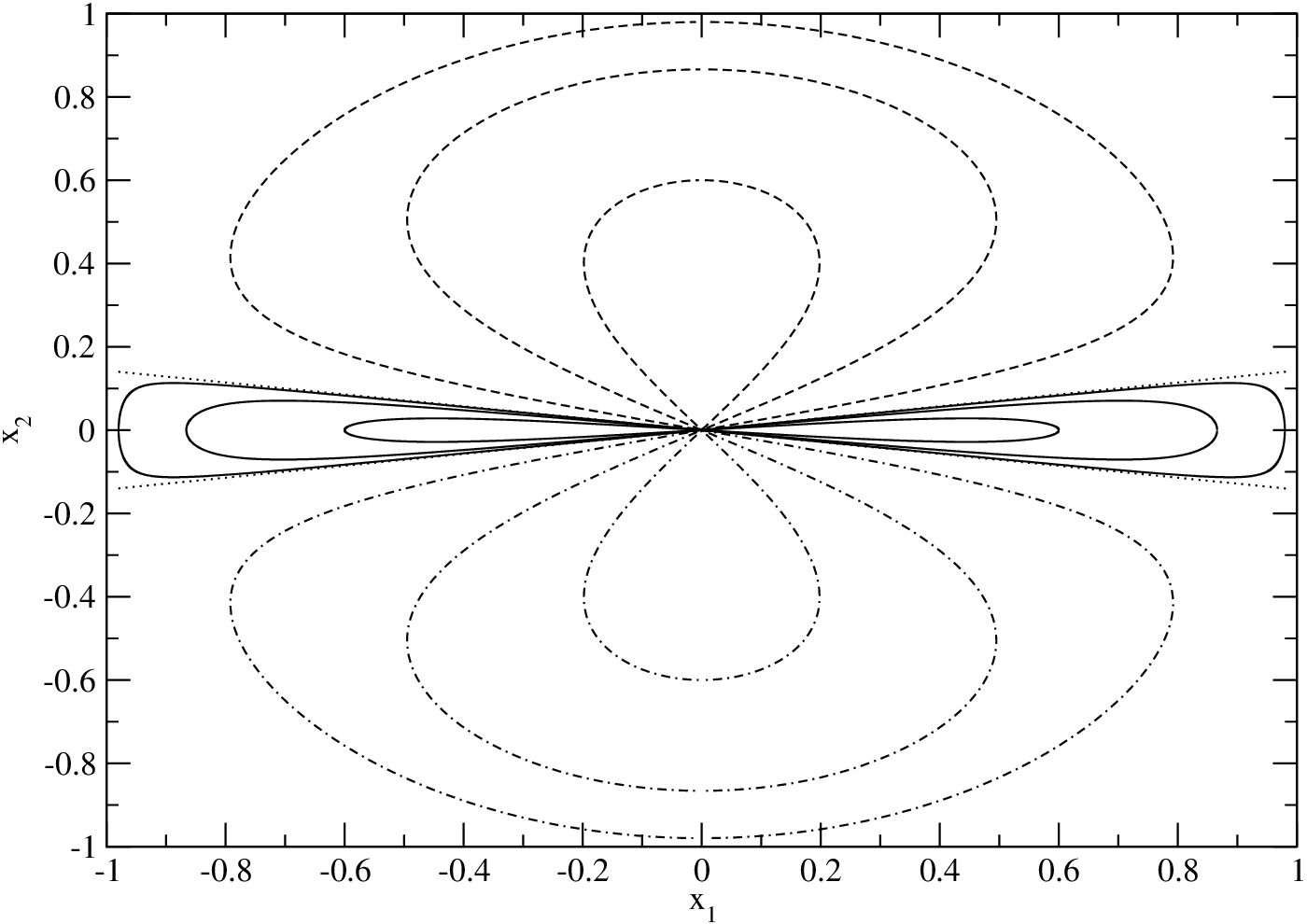}
\caption{ Similar to Fig. 2 but now $\kappa= 1.9$ is just below the
critical value $\kappa_{\rm crit}=2$. All other parameters are the
same as in Fig. 2. The shape of the curves of the different types is
similar to the case $\kappa=0$, but the lines separating the regions
of different evolution have smaller inclination with respect to the
axis $x_{1}$ and the librating solutions occupy more space on the
plane $(x_{1}, x_{2})$.}
\end{figure} 

\begin{figure}
\vspace{8cm}\includegraphics{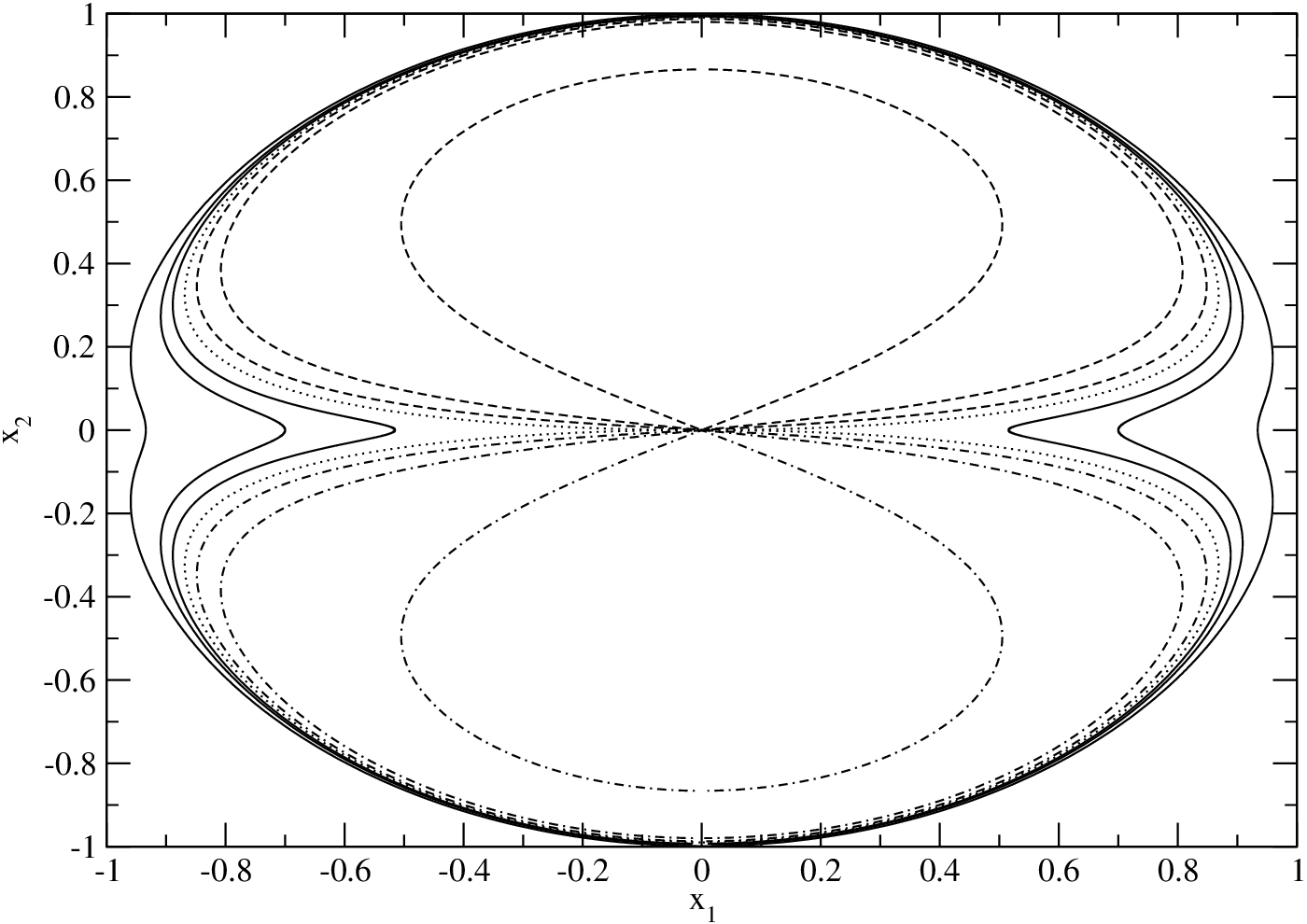}
\caption{The case of $\kappa =2.1 > \kappa_{\rm crit}$ is shown. The
value of $\Lambda$ is the same as in the previous cases.  The shape of
the curves of $(r)$ type is drastically different from the cases
$\kappa=0$ and $\kappa=1.9$. Both rotating and librating curves have
the minimal values of the eccentricity when crossing the line
$x_{1}=0$ ($\omega=\pm \pi/2$).  When this value is smaller (larger)
than the critical value $e_{\rm crit}\approx 0.14$ the curves belong
to the family of $(r)$ type curves ($(l)$ type curves). As follows
from the text for the curves of $(r)$ type we define the value of
eccentricity corresponding to $\omega=\pm p/2$ as $e_{1}$.  We show,
respectively, the curves corresponding to $e_{1}=0.05$, $0.1$, $0.12$
(the solid curves) and the curves corresponding to $e_{-}=0.16$, $0.2$
and $0.5$ (the dashed and dot dashed curves). The dotted `critical'
curve represents the boundary between the regions on the plane
$(x_{1},x_{2})$ where the motion of librating and rotating type is
possible.}
\end{figure} 

\begin{figure}
\vspace{8cm}\includegraphics{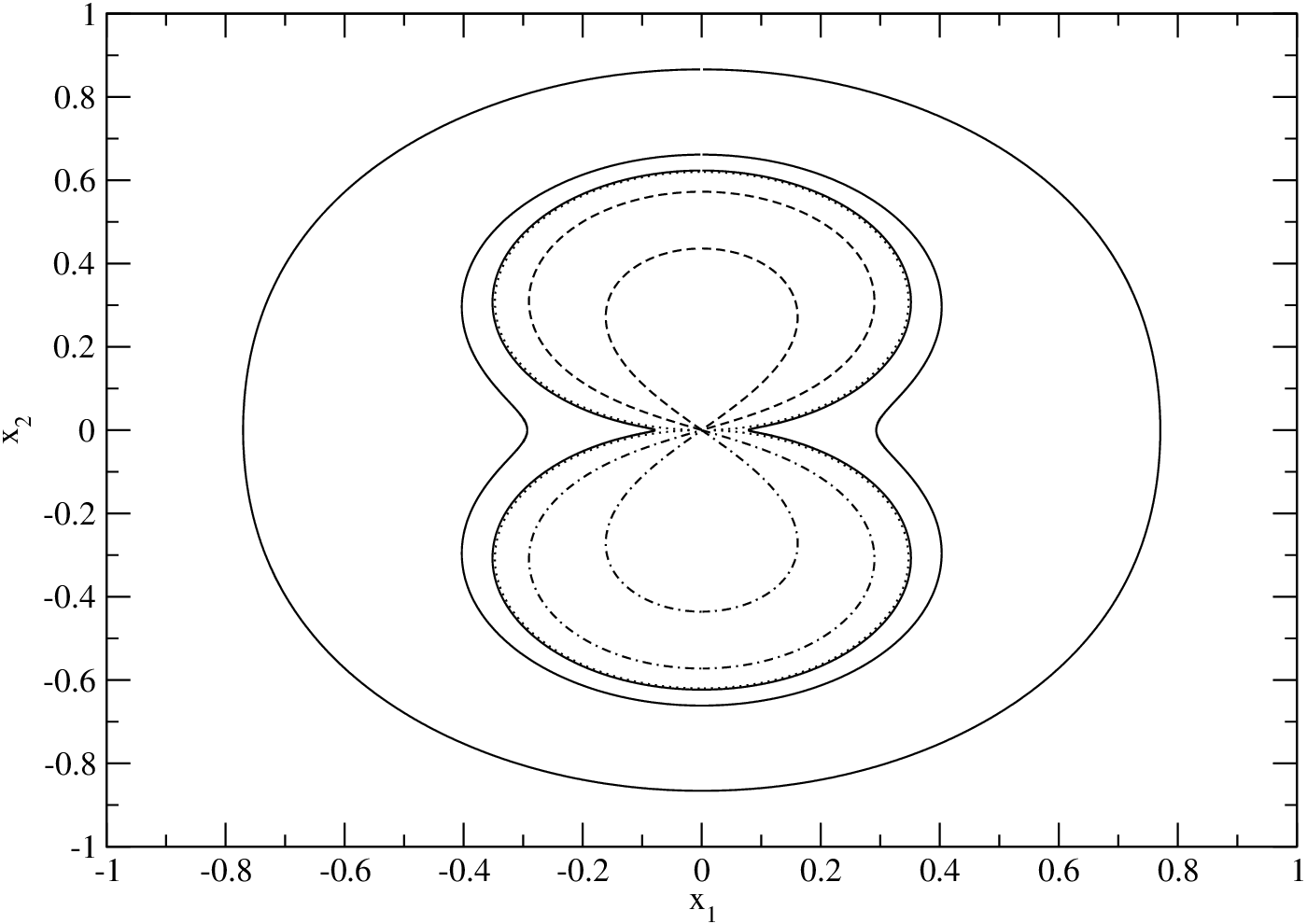}
\caption{Similar to Fig. 4 but corresponding to $\kappa=10$. In this
case $e_{\rm crit}\approx 0.784$ and we show the curves corresponding
to $e_{1}=0.5$, $0.75$, $0.782$ (the solid curves) and $e_{-}=0.82$
and $0.9$ (the dashed and dot dashed curves). Note that apart from a
small region near the origin of coordinates the dotted curve
representing the boundary between the corresponding regions almost
coincide with the curve corresponding to $e_{1}=0.782$.}
\end{figure}

\subsection{The structure of the phase space}

It is convenient to represent the solutions of the different types on
the plane $(x_{1}=\epsilon \cos (\omega )$, $x_{2}=\epsilon \sin
(\omega))$. Since $\epsilon=\sqrt{(1-e^{2})} < 1$, all solutions are
situated within the unit circle.  Provided that the integral $\Lambda
$ is fixed the curves corresponding to different solutions belong to
the same two-dimensional dynamical system and do not cross each other,
and we use the small $\Lambda=1\cdot 10^{-3}$ hereafter.  We show the
solutions in Figs.~2--5. Fig.~2 and Fig.~3 represent the case of
$\kappa < \kappa_{\rm crit}=2$, specifically $\kappa=0$ and $\kappa=1.9$.
The curves representing the solutions of $(r)$ type have a `propeller'
like shape, and the curves representing the solutions of $(l)$ type
have a `bulb' like shape with `cones' directed toward the origin of
the coordinate system. When the eccentricity is not very large and the
integral $\Lambda $ is small, the curves separating the regions in the
plane $(x_{1},x_{2})$, which correspond to the solutions of different
types, are approximately two straight lines. Setting the value of
$e_{0}$ to zero in equation (21), substituting the result in equation
(18) and neglecting the value of $\Lambda $ in denominator of the
result we can find the inclination angles, $\omega_{\rm crit}$, of
these lines
$$\sin{(\omega_{\rm crit})}=\pm \sqrt{{2-\kappa \over 5}}. \eqno 33$$ 
From equation (33) it follows that when $\kappa \rightarrow \kappa_{\rm crit}$, 
$\omega_{\rm crit}\rightarrow 0$.

Fig.~4 and Fig.~5 represent the case of $\kappa > \kappa_{\rm
crit}$. As we discussed above, in this case, the behaviour of the
solutions of $(r)$ type changes drastically. Contrary to the case
$\kappa < \kappa_{\rm crit}$, the solutions have minimal eccentricity
at $\omega = \pm \pi/2$ and maximal eccentricity at $\omega=0$, $\pi$.
Also, the angle $\omega $ is rotating in the opposite negative
direction compared to the previous case.  The behaviour of solutions
changes drastically and sharply when $\kappa$ exceeds $\kappa_{\rm
crit}$.  For example, in Fig.~4 we show the case of $\kappa =2.1$
which is just above the critical value of $\kappa=2$. One can see from
this figure and Fig.~5 that only the curves of $(r)$ type having the
maximal values of eccentricity very close to the `critical' value
$e_{\rm crit}=\sqrt{{\kappa-2\over \kappa +3}}$ can have a small value
of the minimal eccentricity. Since we are interested only in the
solutions which can have a very large value of the eccentricity the
solutions of $(r)$ type are not interesting for us when $\kappa >
2$. On the other hand the librating solutions can produce a very large
eccentricity, and therefore they are relevant to our problem even when
$\kappa \gg 1$ (see Fig.~5). The solutions of both types are separated
by a `critical' curve which can be obtained from equations (18) and
(26). For that we set $e_{-}=e_{\rm crit}$ in equation (26) and
substitute the result in equation (18). We have
$$\sin(\omega_{\rm crit})=\pm \sqrt{{\kappa -2\over 5}}{\epsilon \over e}. \eqno 34$$
From equation (34) it follows that when $\kappa$ increases the size of
the regions on the plane $(x_{1}, x_{2})$ corresponding to the
librating solutions decreases. Equation (32) also shows that the
solution (34) formally has infinite period.

\section{The supply rate of the stars to the loss cone} 

Now we would like to estimate the number of stars supplied to the loss
cone as a function of time assuming that the initial distribution
function is given by equation (9).  To begin with we consider the
Keplerian case, $\kappa=0$.  We also assume that the Kozai effect is
solely responsible for the change of the angular momentum of the
stars. As we have mentioned above, the $z$-component of the angular
momentum is approximately conserved in this case, and the number of
stars per unit semi-major axis that can be supplied to the loss cone
and tidally disrupted due to this effect can be estimated as the
number of stars (per unit semi-major axis), $N^{c}_{0}(a)$ with $L_z$
smaller than the size of the loss cone $L_{T}$.  After integration
over $L_{z} < L_{T}$ and $L$ of the distribution (9) we obtain
$$N^{c}_{0}(a)=\int^{L_{a}}_{0}dL \int_{-L_{T}}^{L_{T}}dL_{z} N_{0}(a,L,L_{z})
=2N_{0}(a){l_{T}\over \sqrt {\tilde a}}, \eqno 35$$
where $N_{0}(a)$ is given by equation (7).

Let us estimate the time dependence of the number of stars supplied to
the loss cone.  After the moment of time $t=0$ the number of stars
with $L_{z} < L_{T}$ per unit of $a$, $N^{c}(a)$ is decreasing with
time, and the rate of this decrease can be estimated as
$${d N^{c}(a)\over dt}= -{N^{c}(a)\over P^{0}_{K}}, \eqno 36$$ 
where $P^{0}_{K}$ is given by equations (24) and (29) for the
solutions of (r) and (l) type, respectively. As follows from this
equation $P_{K}$ depends on the eccentricity $e$, and therefore,
strictly speaking, in equation (36) we should divide the distribution
function on $P^{0}_{K}$ before integration of the distribution
function over the absolute value of angular momentum. However this
dependence is logarithmic, say $P^{0}_{K}(e=0.5) \approx 1.48T_{K}$
and $P^{0}_{K}(e=0.9) \approx T_{K}$.  Therefore, for simplicity, we
neglect this dependence hereafter and set $P_{K}=T_{K}$ from now
on. Integrating equation (36) we obtain
$$N^{c}(a)=N^{c}_{0}(a)e^{-({\tilde a}^{3/2}\tau )}, \eqno 37$$
where we use equation (14), $\tau={3q\over 4}\Omega_{b}t $, and
$\Omega_{b}=\sqrt {{GM\over D^{3}}}$. The number of stars $N^{t}(a)$
supplied to the tidal loss cone and tidally disrupted follows from
equations (35) and (37):
$$N^{t}(a)=N^{c}_{0}(a)-N^{c}(a)=2N_{0}(a){l_{T}\over
\sqrt {\tilde a}}(1-e^{-({\tilde a}^{3/2}\tau )}). \eqno 38$$ 

The total number $N^{t}(t)$ of stars supplied to the loss cone is
obtained by integration of equation (38) over $a$. The integration is
straightforward, with the result
$$N^{t}(t)=\int^{a_{\rm max}}_{0}daN^{t}(a)= {2l_{T}\over (1-p)}
C_{a}D^{(3/2-p)}{\tilde a_{\rm max}}^{(1-p)}(1- {2(1-p)\over
3}x^{-2/3(1-p)}\gamma (2/3(1-p); x)), \eqno 39$$
where $x={\tilde a_{\rm max}}^{3/2}\tau$, and $\gamma (\beta, x)$ is the
incomplete gamma function, $a_{\rm max}$ is the upper limit of integration
and we set the lower limit to zero. As we have discussed above we
assume that $\tilde a_{\rm max} \approx 1/2$ for the inner
problem\footnote{Note, that the quadrupole approximation used in our
analysis is, strictly speaking, invalid in the case of rather large
$\tilde a_{\rm max} =1/2$. However, we assume that the estimates based on
this assumption are correct in order of magnitude.}.  Note, that to
obtain equation (39) we use the explicit form of the initial
distribution function over $a$ given by equation (7).

It is very important to point out that $N^{t}(t)$ depends linearly on
the small parameter $l_{T}$. The fact that the dependence is linear
leads to a significant increase of tidally disrupted stars in the
presence of the secondary black hole (see Section 6). In contrast to
this result the usual estimate of the tidal disruption rate due to
two-body star-star gravitational scattering, in the so-called `full'
loss cone approximation (e.g., Hills 1975, Frank \& Rees 1976) leads
to a quadratic dependence. In the last case, the tidal disruption rate
is proportional to the probability to find a star inside a circle in
the angular momentum space of radius $L_{T}$, which is proportional to
the area of this circle $\pi L_{T}^{2}$. As we have mentioned above,
in the presence of the secondary black hole only the $z$ component of
angular momentum is conserved, while total angular momentum varies due
to the Kozai effect and can be evolved to a very small value $L <
L_{T}$. Hence, in our case the tidal disruption rate is proportional
to probability of finding a star with $L_{z} < L_{T}$. Note that in
this respect our problem is similar to the problem of tidal disruption
in a non-spherical galaxy where there is a tidal loss `wedge' instead
of the tidal loss cone, and the tidal disruption rate is enhanced
(e.g., Magorrian \& Tremaine 1999).

For comparison of our theory with results of numerical computations it
is more convenient to use the probability $\prob(t)$ for a star to
find itself in the tidal loss cone. This can be obtained dividing the
expression (39) by the total initial number of stars with semi-major
axes smaller than $a_{\rm max}$, $N_{\rm tot}$. This number directly
follows from equation (7)
$$N_{\rm tot}={C_{a}\over (3/2-p)}a_{\rm max}^{3/2-p}. \eqno 40$$
Using equations (36) and (40) we obtain
$$\prob(t)={(3-2p)\over (1-p)}{l_{T}\over \sqrt {\tilde a_{\rm max}}}(1-
{2(1-p)\over 3}x^{-2/3(1-p)}\gamma (2/3(1-p); x)). \eqno 41$$   
  
The estimates shown above are approximately valid only for
sufficiently large values of the loss cone $l_{T}$. The results of
numerical calculations (see the next Section) show that when $l_{T}$
is sufficiently small equations (39) and (41) significantly
underestimate the number of stars supplied to the loss cone. To
explain this effect let us consider the evolution of angular momentum
on time scales of the order of the binary orbital period $P_{b}=2\pi
\Omega_{b}^{-1}$, which is much smaller than the Kozai time scale
$T_{K}$. It turns out that the angular momentum oscillates on the time
scale $\sim P_{b}$ around an averaged value determined by the secular
evolution on the time scale $\sim T_{K}$. The largest amplitude of
these oscillations is reached near the maximal value of the averaged
eccentricity $e_{\rm max}$ provided by the Kozai effect, and,
respectively, near the minimal value of the angular momentum (see
Appendix 1).  The maximal amplitude, $\Delta L$ is estimated in
Appendix B as
$$\Delta l ={15\over 8}\sqrt {{3\over 5}}q{\tilde a}^{2}, \eqno 42$$
where $\Delta l = \Delta L/\sqrt{GMD}$, and we take into account the
fact that when $\kappa=0$ both rotating and librating solutions have
$\cos (i_{\rm min})\approx \sqrt{3\over 5}$ (see equations (20) and (29)).
It follows from equation (42) that when the size of the loss cone is
very small, $l_{T} \ll q$, the maximal value of $L_z$ of stars that
can be supplied to the loss cone is of the order of $\Delta l \gg
l_{T}$.  Accordingly, the number of stars supplied in the loss cone is
determined by $\Delta l$:
$$N^{c}_{0}(a)=\int^{L_{a}}_{0}dL \int_{-\Delta L}^{\Delta L}dL_{z}
N_{0}(a,L,L_{z})=
2N_{0}(a){\Delta l\over \sqrt {\tilde a}}=
{15\over 4}\sqrt{3\over 5} q C_{a}D^{1/2-p}{\tilde a}^{2-p}, \eqno 43$$
where we use equations (7), (9) and (42).  

Similar to the case discussed above the number of stars per unit
interval of $a$ with $L_{z} < \Delta L$, $N^{c}(a)$ decreases with
time. It is important to note that in the case of the small tidal loss
cone the probability of a star with $L_{z} < \Delta L$ to obtain the
absolute value of angular momentum smaller than $L_{T}$ during one
Kozai cycle is of the order of $L_{T}/\Delta L \ll 1$. Accordingly,
the rate of decrease of the number of stars with $L_{z} < \Delta L$
(per unit of semi-major axis) can be estimated as
$${d N^{c}(a)\over dt}\approx -\left({L_{T}\over \Delta L}\right)
{N^{c}(a)\over T_{K}}=
-{2 l_{T}\over \sqrt {15 \tilde a}} N^{c}(a)\Omega_{b}. \eqno 44$$ 
Integrating equation (44) we obtain
$$N^{c}(a)=N^{c}_{0}(a)e^{-(\tau_{1}/ \sqrt {\tilde a})}, \eqno 45$$
where $\tau_{1}={2 l_{T}\over \sqrt 15}\Omega_{b}t$, and the number of
stars supplied to the loss cone is
$$N^{t}(a)=N^{c}_{0}(a)-N^{a}(a)=
N^{c}_{0}(a)(1-e^{-(\tau_{1}/ \sqrt {\tilde a})}). \eqno 46$$
The total number $N^{t}(t)$ of stars supplied to the loss cone is
obtained again by integration of equation (46) over $a$ with the
result
$$N^{t}=\int^{a_{\rm max}}_{0}da N^{t}(a)=
{15\over 4}\sqrt {3\over 5} q {{\tilde a_{\rm max}}^{3-p}\over (3-p)}C_{a}D^{3/2-p}
(1-I), \eqno 47$$
where
$$I={{3-p}\over {\tilde a_{\rm max}}^{3-p}}
\int^{\tilde a_{\rm max}}_{0}d\tilde a \, {\tilde a}^{2-p}
e^{-(\tau_{1}/ \sqrt {\tilde a})}, \eqno 48$$
and the probability for a star to find itself in the tidal loss cone,
$\prob_{1}=N^{t}/N_{\rm tot}$ is
$$\prob_{1}={15\over 8}\sqrt {3\over 5} {(3-2p)\over (3-p)}q{\tilde a_{\rm max}}^{3/2} 
(1-I). \eqno 49$$
The integral (48) can be estimated in terms of elementary functions in
two limiting cases. Assuming that the argument in the exponent in (48)
is sufficiently small, $\tau_{1} < \sqrt {\tilde a_{\rm max}}$, we can set
$\tilde a=\tilde a_{\rm max}$ in the exponent. In this case we have
$$I\approx e^{-(\tau_{1}/ \sqrt {\tilde a_{\rm max}})}, \eqno 50$$
and in the opposite case of a large time $\tau_{1} > \sqrt {\tilde
a_{\rm max}}$, the integral is estimated to be
$$I\approx 2(3-p){\sqrt { \tilde a_{\rm max}} \over \tau_{1}}e^{-(\tau_{1}/ \sqrt {\tilde a_{\rm max}})}. \eqno 51$$

Taking into account the effect of additional oscillations of the
angular momentum on the time scale of the binary orbital period
significantly improves agreement between our theory and numerical
calculations (see the next Section).

\subsection{The case $\kappa \ne 0$}

When $\kappa \ne 0$ the estimates of the number of stars supplied to
the tidal loss cone must be modified. However, as we discussed in the
previous Section, when $\kappa < \kappa_{\rm crit}=2$ the secular
evolution of the orbital elements of the stars is very similar to the
case $\kappa=0$, and therefore we do not discuss this case later on,
assuming that our expressions (39), (41) and (47), (49) give a correct
order of magnitude estimate. When $\kappa > 2$ only the librating
solutions of the secular equations (10--13) could produce a
significant change of the eccentricity.  Therefore, only the stars
having orbital elements corresponding to the secular evolution of (l)
type can be tidally disrupted provided that the $L_z$ of these stars
is sufficiently small. As follows from equation (34) the boundary
between the solutions of (l) and (r) types can be defined as a closed
`critical' curve in the plane $(\epsilon, \omega)$, and therefore, in
order to estimate the number of stars going to the loss cone, we
should introduce the initial distribution function of stars over the
semi-major axes, the quantities $\epsilon =\sqrt{1-e^{2}}$, $L_z$ and
the argument of pericentre $\omega$, $N_{0}(a,\epsilon, L_{z},
\omega)$. Taking into account that the distribution of stars is
assumed to be the uniform over $\omega $ and that $\epsilon =L/L_{a}$,
the distribution function $N_{0}(a,\epsilon, L_{z}, \omega)$ can be
easily obtained from equation (9):
$$N_{0}(a,\epsilon, L_{z}, \omega)={N_{0}(a)\over 2\pi L_{a}}. \eqno 52$$ 
We find the number of stars (per unit semi-major axis) that can be
supplied to the tidal loss cone, $N^{c}_{0,\kappa}$, integrating the
distribution function (52) over the surface under the `critical' curve
(34), and also integrating over $L_z$:
$$N^{c}_{0,\kappa}(a)=\int^{L_{*}}_{-L_{*}}dL_{z}
\int_{\!_{\!_{\!_S}}} \! \! \int dS\
N_{0}(a,\epsilon, L_{z}, \omega)=
{4\over \pi}\arcsin\left(\sqrt{5\over \kappa +3}\right)
{l_{*}\over \sqrt {\tilde a}}N_{0}(a), \eqno 53$$  
where $\int_{\!_{\!_{\!_S}}} \! \! \int dS\ (...)$ implies integration
over the surface in the plane $(\epsilon, \omega)$ within the
`critical' curve (34), and $l_{*}$ is the absolute value (in units of
$\sqrt {GMD}$) of the $z$ component of angular momentum of the stars
supplied to the loss cone. As we discussed above when $l_{T} \gg q$ we
have $l_{*} \approx l_{T}$ and in the opposite limit $l_{*}\approx
\Delta l$ where $\Delta l$ is given by equation (42).
  
As follows from equations (30) and (32) the period of Kozai cycles
$P_{K}$ depends on $\kappa$, and on the maximal and minimal
eccentricities $e_{\pm}$. Also, the period is formally infinite for
the solution corresponding to the critical curve (34). We are
interested in the case of a high maximal eccentricity $e_{\rm max} \approx
1$ and the minimal eccentricity close to the critical eccentricity
$e_{\rm crit}$. Therefore, we use the approximate value of $P_{K}$
given by equation (32).  Note that this value formally depends on the
eccentricity $e_{-}$ through the quantity $\delta=(e_{-}-e_{\rm
crit})/e_{\rm crit}$.  However, since $\delta $ enters the expression
(32) only in the logarithmic factor we expect that the use of a
definite value of $\delta$ is irrelevant for our order of magnitude
estimates, and for definiteness we set its value to the `typical'
value $\delta_{*}=\sqrt{5\over \kappa +3}$.  We have:
$$P_{K}\approx  {2\ln {({2\sqrt{2})}}
\over\sqrt{5(\kappa-2)}}T_{K}\approx
{2\over \sqrt{5(\kappa-2)}}T_{K}.  \eqno 54$$   

It is very important to note that both the number of stars supplied to
the loss cone $N^{c}_{0,\kappa}(a)$ and the period of Kozai cycles
$P^{\kappa}_{K}$ are proportional to the same quantities corresponding
to the case $\kappa=0$: $P_{K}=C_{1}T_{K}$, and
$N^{c}_{0,\kappa}(a)=C_{2}N^{c}_{0}(a)$, where $C_{1}$ is determined
by equation (54), and $C_{2}$ follows from equations (35), (43) and
(53):
$$C_{2}={4\over \pi}\arcsin\left(\sqrt {5\over \kappa +3}\right). \eqno  55 $$ 
That means that the results obtained above for the case $\kappa =0$
can be used when calculating the number of stars supplied to the loss
cone for a non-zero value of $\kappa $. We should only multiply the
expressions (39), (40) and (47), (49) by the correction factor $C_{2}$
and also make an appropriate change of the time variables entering
these expression with the help of equation (54).  When $l_{T} \gg q$ we
obtain from equations (41), (54) and (55) the expression for the
probability for a star to find itself in the loss cone,
$\prob_{\kappa}(t)$:
$$\prob_{\kappa}(t)=C_{2}\prob(t)={4(3-2p)\over\pi (1-p)}
\arcsin\left({\sqrt {5\over \kappa +3}}\right)
{l_{T}\over \sqrt {\tilde a_{\rm max}}}
\left(1-{2(1-p)\over 3 x^{2/3(1-p)}}\gamma ({2\over 3}(1-p);
x_{\kappa})\right), \eqno 56$$
where
$$x_{\kappa}=x/C_{1}={3\over 8}\sqrt{5(\kappa -2)}
{\tilde a_{\rm max}}^{3/2}q\Omega_{b}t. \eqno 57$$
In the opposite case $l_{T} \ll q$ we use equations (49), (54) and
(55)
$$\prob_{1,\kappa}(t)=C_{2}\prob_{1}(t)=
{15(3-2p)\over 2\pi (3-p)}\arcsin\left(\sqrt{5\over \kappa +3}\right)
q{\tilde a_{\rm max}}^{3/2} (1-I(\tau_{1,\kappa})), \eqno 58$$
where
$$\tau_{1,\kappa}=\sqrt{3\over 5}\tau_{1}/C_{1}=
\sqrt{\kappa -2\over 5}l_{T}\Omega_{b}t, \eqno 59$$
and the quantity $I$ is given by equation (48).  Note that we discard
the factor $\sqrt{3\over 5}$ in equation (42) and accordingly in
equations (44)-(49) to obtain equations (58) and (59). This accounts
for the coefficient $\sqrt{3\over 5}$ in equation (59).  As follows
from equation (25), for the eccentricities $e_{-} \approx e_{\rm
crit}$, $\sin (i_{\rm min})\approx 0$ and accordingly $\cos
(i_{\rm min})\approx 1$. This is different from the case of $\kappa =0$
where $\cos (i_{\rm min})\approx \sqrt{3\over 5}$.

\subsection{Characteristic rates and time scales}

In general the expressions for the number of stars supplied to the
loss cone given by equations (39), (41), (47), (49) and (56), (58),
are rather complicated functions of time. They also involve many
parameters determined by the properties of a particular stellar
cluster and a binary black hole. These complicated expressions are
rather difficult to use. Also, in fact, they are not needed for rough
estimates of the tidal disruptions rate in the presence of the black
hole binary.  Here, we would like to obtain much simpler order of
magnitude expressions which will be used in Section 6 for these
estimates.  For simplicity, we also assume that the mass of the
stellar cluster is much larger than the mass of the secondary black
hole, and accordingly, formally consider the case of high values of
$\kappa \gg 1$.

In order to obtain the total probabilities for a star to be tidally
disrupted let us consider the limit $t\rightarrow \infty$ in equations
(56) and (58). In this limit, the second terms in the brackets in (56)
and (58) are equal to zero and we have
$$\prob_{\rm tot}=\prob_{\kappa}(t\rightarrow \infty)\approx
{4\sqrt 5\over \pi}{(3-2p)\over (1-p)}
{l_{T}\over \sqrt{\tilde a_{\rm max} \kappa}} \eqno 60$$
from equation (56), and
$$\prob_{tot,1}=\prob_{1,\kappa}(t\rightarrow \infty)\approx
{15\sqrt 5\over 2\pi} {(3-2p)\over (3-p)}
{{\tilde a_{\rm max}}^{3/2}\over \sqrt \kappa}q,
\eqno 61$$
from equation (58). The characteristic tidal disruption rate
$\dot\prob$ per star can be obtained from equations (56) and (58) by
differentiation these equations over the time and setting $t=0$ in the
resulting expressions. We have
$$\dot \prob=\dot \prob_{\kappa}(t=0)=
\dot \prob_{1,\kappa}(t=0)={15\over \pi}{(3-2p)\over (5-2p)}
\tilde a_{\rm max}ql_{T}\Omega_{b}^{-1}. \eqno 62$$
Let us note that the quantity $\dot \prob$ does not depend on
$\kappa$.  The characteristic time scales associated with the process
of tidal disruption, $T_{*}$ and $T_{*,1}$ can be obtained from
equations (60), (61) and (62). When $l_{T} \gg q$ we have
$$T_{*}=\prob_{\rm tot}/\dot \prob=
{4\over 3\sqrt 5}{(5-2p)\over (1-p)}
{{\tilde a_{\rm max}}^{-3/2}\over \sqrt \kappa}{(q\Omega_{b})}^{-1}, \eqno 63$$
and in the opposite case $l_{T} \ll q$ we have
$$T_{*,1}=\prob_{tot,1}/\dot \prob=
{\sqrt 5\over 2}{(5-2p)\over (3-p)}
{{\tilde a_{\rm max}}^{1/2}\over \sqrt \kappa}{(l_{T}\Omega_{b})}^{-1}. \eqno 64$$
As we discussed in Section 2, the typical values of the parameter $p$
are close to zero. Therefore, we can further simplify expressions
(60)-(63) setting $p=0$ and $\tilde a_{\rm max}=1/2$ and expressing
$\kappa$ in terms of the ratio of the total mass of the stars to the
mass of the secondary, $M_{st}/m$ with help of equations (15) and
(16):
$$\kappa={8\sqrt 2\over 3\pi}{M_{st}(a_{\rm max})\over m}=
{4\over 3\pi}{M_{st}(D)\over m} \approx 0.4{M_{st}(D)\over m}, \eqno 65$$
where we take into account that when $p=0$, $M_{st}(a)\propto a^{3/2}$
(see equation 4).

We have
$$\dot \prob ={9\over 2\pi} ql_{T}\Omega_{b}^{-1}\approx
1.4 ql_{T}\Omega_{b}  \eqno 66$$
from equation (62),
$$\prob_{\rm tot}=6\sqrt{30\over \pi}\sqrt {m\over M_{st}}l_{T}\approx
18.5\sqrt {m\over M_{st}}l_{T}  , \eqno 67$$
$$T_{*}=4\sqrt{10\pi\over 3}\sqrt {m\over M_{st}}{(q\Omega_{b})}^{-1}\approx
13\sqrt {m\over M_{st}}{(q\Omega_{b})}^{-1}, \eqno 68$$
from equations (60), (63), and
$$\prob_{tot,1}={15\over 8}\sqrt{15\over 2\pi}
\sqrt {m\over M_{st}}q\approx 2.9\sqrt {m\over M_{st}}q, \eqno 69$$
$$T_{*,1}={5\over 4}\sqrt{5\pi \over 6}
\sqrt {m\over M_{st}}{(l_{T}\Omega_{b})}^{-1}\approx
2\sqrt {m\over M_{st}}{(l_{T}\Omega_{b})}^{-1}, \eqno 70$$
from equations (61) and (64). Equations (66--70) are used in Section 6
for estimates of the tidal disruption rate in astrophysical
systems. Although equations (66--70) have been formally obtained in
the limit $\kappa \gg 1$ they can also be used for order of magnitude
estimates even when $\kappa \sim 1$.

\section{Numerical calculations}

Now we would like to discuss our numerical calculations of the supply
rate and compare them to the analytical theory developed above. In our
numerical work we neglect the influence of the stellar cluster on the
orbits of the stars, and accordingly set $\kappa =0$ reserving the
general case $\kappa \ne 0$ for a future investigation.  We consider
evolution of a large number of particles in the Keplerian
gravitational field of the binary for a sufficiently long time,
assuming that all particles which have the absolute value of angular
momentum smaller than a certain value $l_{T}$ are `tidally disrupted'
by the primary black hole.  The `tidally disrupted' particles are
removed from the calculations.  The initial orbital elements of the
particles are randomly distributed according to the distribution (8)
with $p=0$. The initial semi-major axes of the particles are in the
range $(0.1 < \tilde a < 0.5)$. The calculations depend on two
dimensionless parameters: the mass ratio $q$ and the dimensionless
size of the loss cone $l_{T}$, and we consider the cases $q=1\cdot
10^{-2}$, $3\cdot 10^{-2}$, $1\cdot 10^{-1}$ and $l_{T}=1\cdot
10^{-3}$, $3\cdot 10^{-3}$, $1\cdot 10^{-2}$, $3\cdot 10^{-2}$,
$1\cdot 10^{-1}$, respectively. For given values of $q$ and $l_{T}$ we
evolve approximately $10^{4}-5\cdot 10^{4}$ particles with different
orbital parameters during the time $t <t_{\rm max}={30\over
q}\Omega_{b}^{-1}$.

\begin{figure}
\vspace{8cm}\includegraphics{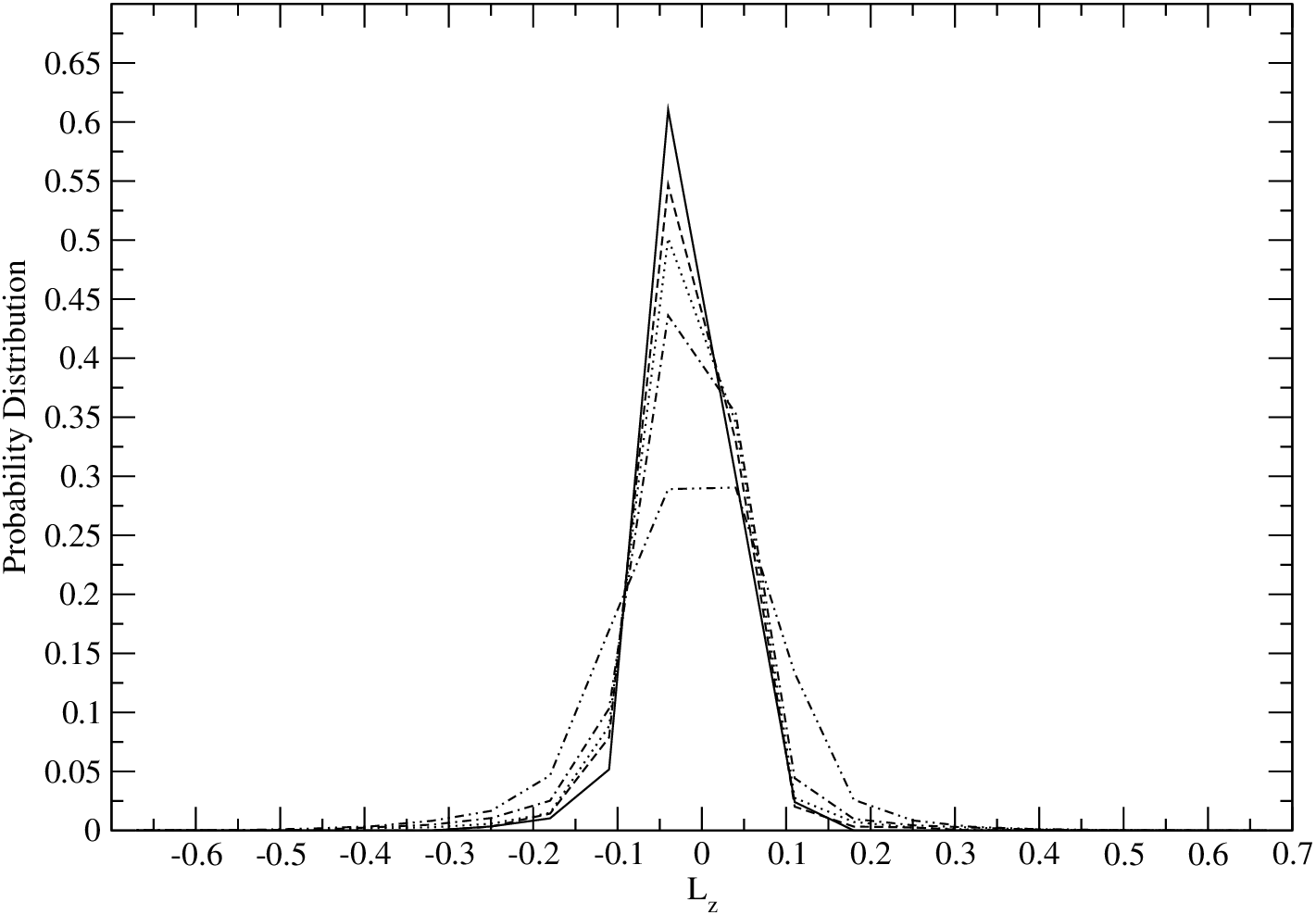}
\caption{The distribution of the `tidally disrupted particles' over
the initial values of the $z$ component of angular momentum (in unit
$\sqrt {GMa}$). The case $q=1$ is shown. The solid, dashed, dotted,
dot-dashed and dot-dot-dashed curves correspond to $l_{T}=1\cdot
10^{-3}$, $3\cdot 10^{-3}$, $1\cdot 10^{-2}$, $3\cdot 10^{-2}$,
$1\cdot 10^{-1}$, respectively. }
\end{figure}

In Fig.~6 we show the distribution of the `tidally disrupted stars'
over the initial values of the $z$ component of angular momentum, for
the case $q=0.1$ and different values of $l_{T}$.  One can see that
that this distribution is strongly peaked at $L_{z}=0$. This strongly
suggests that the effects of the Kozai type are responsible for the
supply of the particles on low angular momentum orbits in our
problem. The spread of the distribution decreases with decrease of
$l_{T}$. However, even for the smallest value of $l_{T}=1\cdot
10^{-3}$ the spread is of the order of $\sim 0.1 \gg l_{T}$. This can
be explained by the additional oscillations of the angular momentum
due to gravitational influence of the secondary on the time scale of
order of the binary orbital period, see equation (42) and Appendix
B. The distributions corresponding to the smaller values of $q$ have a
similar character and are not shown here.

\begin{figure}
\vspace{8cm}\includegraphics{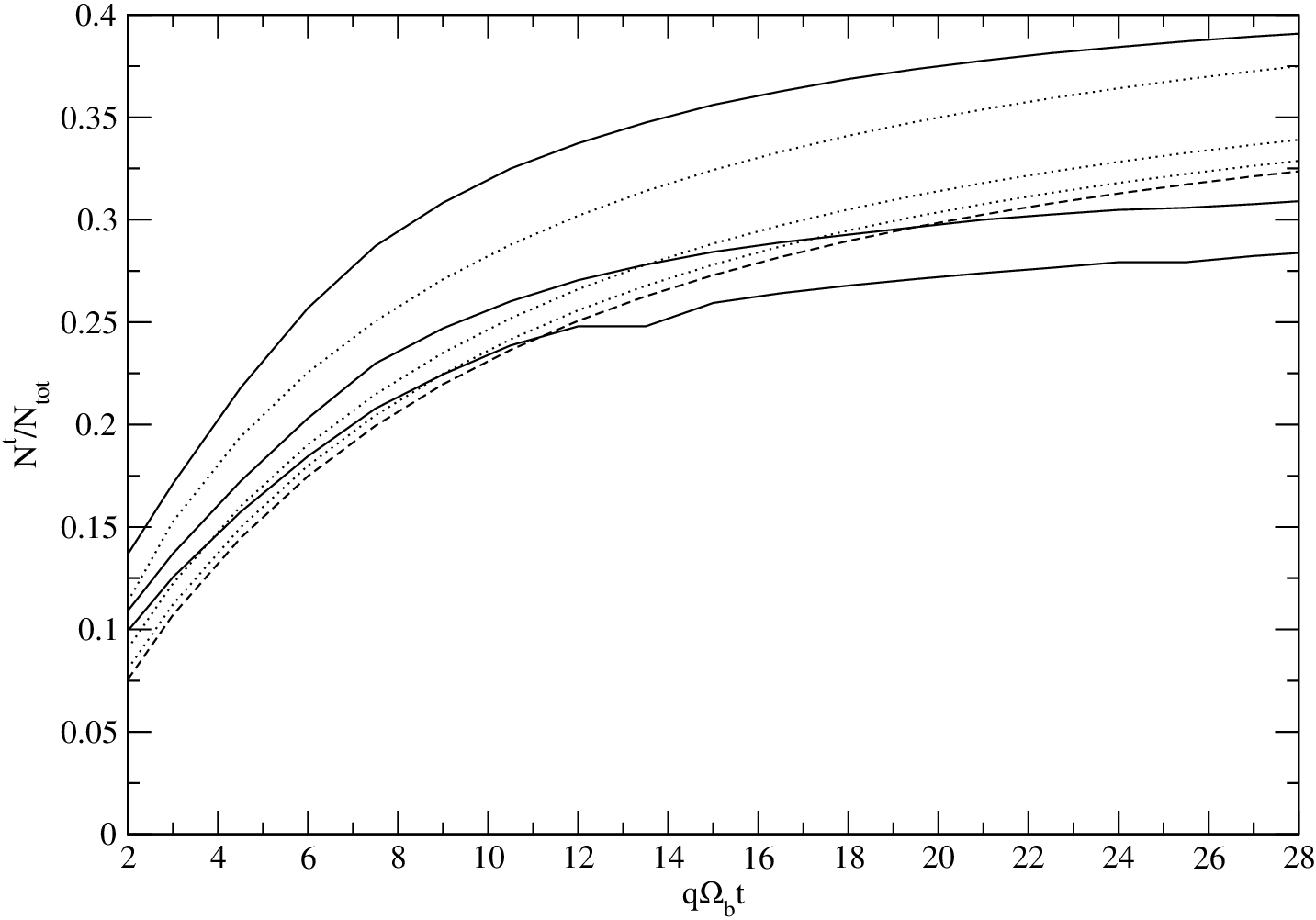}
\caption{The dependence of the number of stars supplied to the loss
cone (in units of the total number of stars participating in the
calculation) as a function of time (in units of
$q^{-1}\Omega_{b}^{-1}$). The size of the loss cone is large:
$l_{T}=0.1$. The three solid curves correspond to $q=1\cdot 10^{-2}$,
$3\cdot 10^{-2}$, $1\cdot 10^{-1}$, with smaller values of ordinates
(at a given moment of time) corresponding to smaller values of
$q$. The dashed curve represents the theoretical solution (41) which
does not depend on $q$ in these time units. The dotted curves
represent the sum of the solutions (41) and (49). The smaller
ordinates correspond again to the smaller values of $q$.}
\end{figure}

In Fig.~7 we show the numerical and theoretical values of probability
for a star to find itself in the loss cone as a function of time, for
a large value of the size of the tidal loss cone. As follows from the
theoretical solution (41), in the simplest approximation this
probability depends on the mass ratio only in the combination
$q\Omega_{b}t$. That means that when the size of the loss cone is
large and fixed, the numerical values of the probability (solid
curves) should be approximately the same being plotted as functions of
the time variable $q\Omega_{b}t$.  In order to check this theoretical
prediction we show the results of the numerical calculations for a
very large and somewhat unrealistic size of the loss cone $l_{T}=0.1$.
As follows from Fig.~7 this theoretical prediction is qualitatively
confirmed in the calculations, although the difference between the
numerical curve corresponding to $q=0.1$ and $q=3\cdot 10^{-2}$ and
$1\cdot 10^{-2}$ is rather large, of the order of $\sim 25$--$30$ per
cent.  The agreement between the theory and numerical calculations can
be improved significantly if we take into account the effect of
oscillation of the angular momentum on the orbital time scale and add
the probability $\prob_{1}$ corresponding to this effect (see equation
49) to the probability $\prob$ determined by the Kozai effect
(equation 25).\footnote{Note that we use the approximate values of the
integral $I$ in the expression for the probability $\prob_{1}$. When
the time variable $\tau_{1} < 6$ we use the expression (33), and when
$\tau_{1} > 6$ we use the expression (34). When $\tau_{1}=6$ both
expressions are equal.} The curves representing the sum $\prob_{\rm
tot}=\prob+\prob_{1}$ are shown as the dotted curves in Fig.~7.  The
curves corresponding to the mass ratios $q=1\cdot 10^{-2}$ and $3\cdot
10^{-2}$ are close to the curve representing $\prob(t)$. The
difference between these curves and those which represent the
numerical calculations is about $\sim 10$--$15$ per cent. The dotted
curve corresponding to $q=0.1$ is rather close to the numerical curve
with the difference of the order of $\sim 6$--$10$ per cent.  Therefore,
when the size of the loss cone is large the use of the total
probability $\prob_{\rm tot}$ instead of $\prob$ improves the
agreement between the theory and the numerical calculations.
  
\begin{figure}
\vspace{8cm}\includegraphics{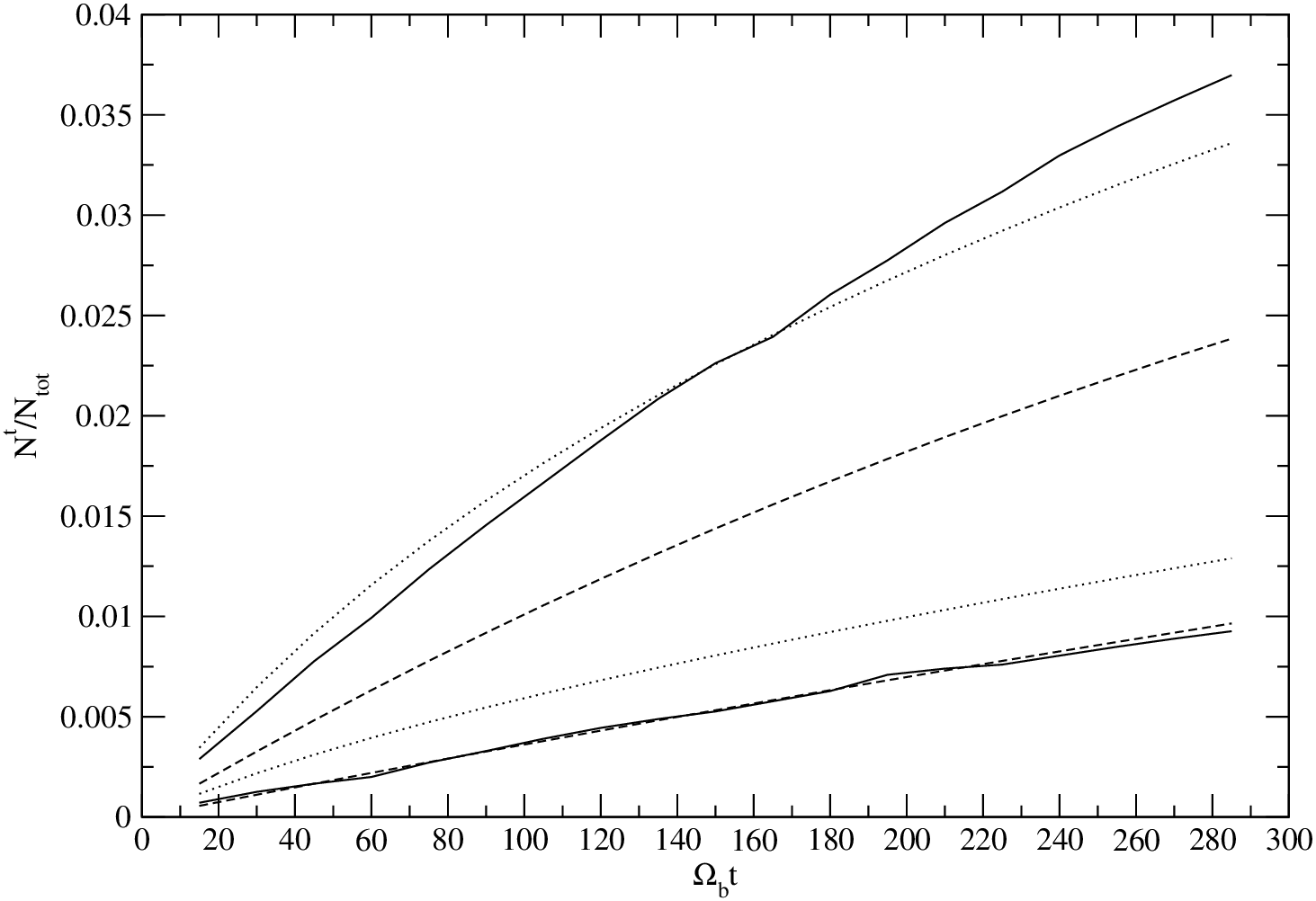}
\caption{The dependence of the number of stars supplied to the loss
cone (in units of the total number of stars) as a function of time (in
units $\Omega_{b}^{-1}$) for $q=0.1$ and the small values of the loss
cone. The upper solid curve corresponds to $l_{T}=3\cdot 10^{-3}$ and
the lower solid curve corresponds to $l_{T}=1\cdot 10^{-3}$.  The
dashed and dotted curves represent the theoretical values of
$\prob_{1}(t)$ and $\prob_{\rm tot}(t)$, respectively. The upper
(lower) dashed and dotted curves correspond to $l_{T}=3\cdot 10^{-3}$
($l_{T}=1\cdot 10^{-3}$).}
\end{figure}
\begin{figure}
\vspace{8cm}\includegraphics{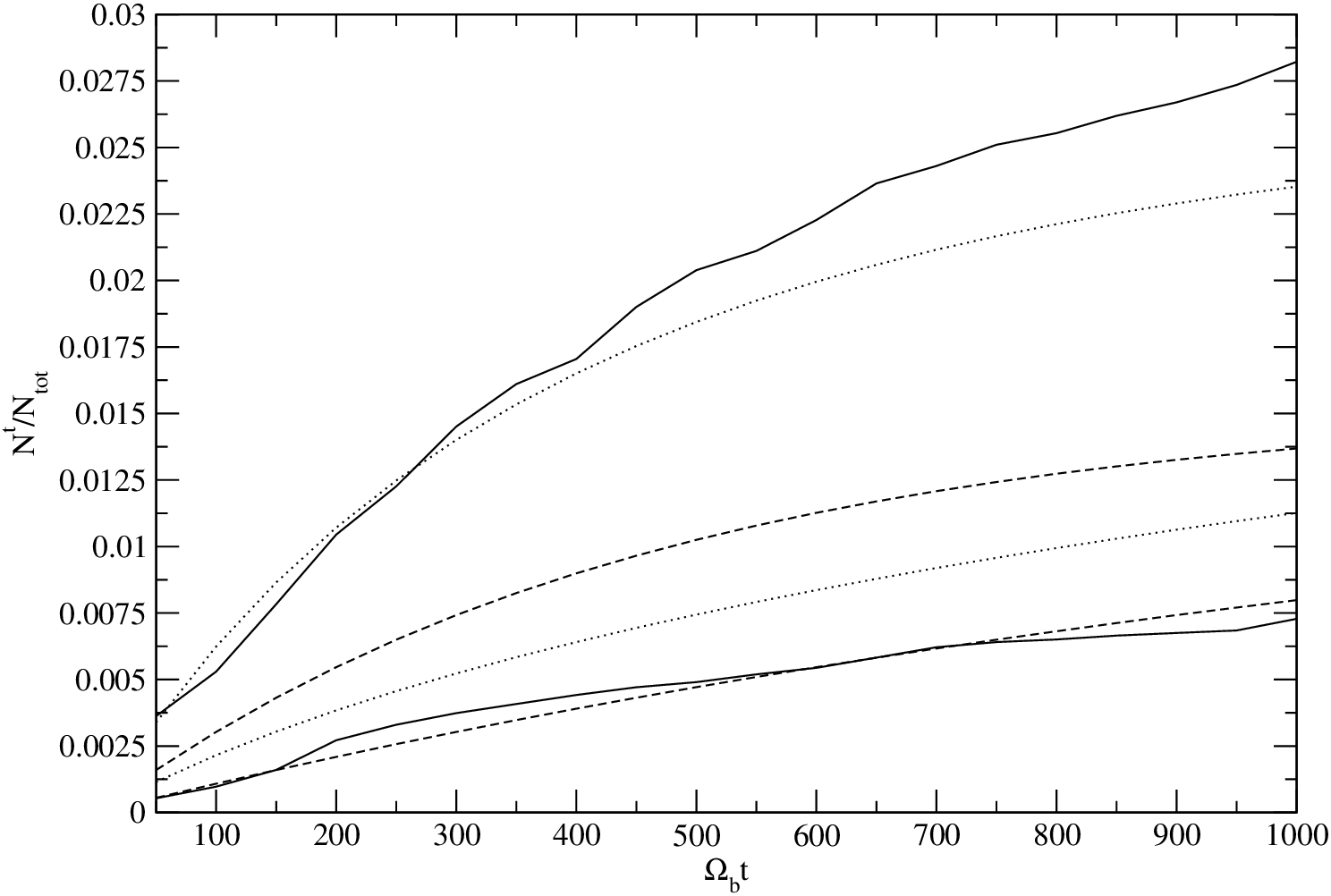}
\caption{Same as Fig.~4 but calculated for $q=3\cdot 10^{-2}$.}
\end{figure}
\begin{figure}
\vspace{8cm}\includegraphics{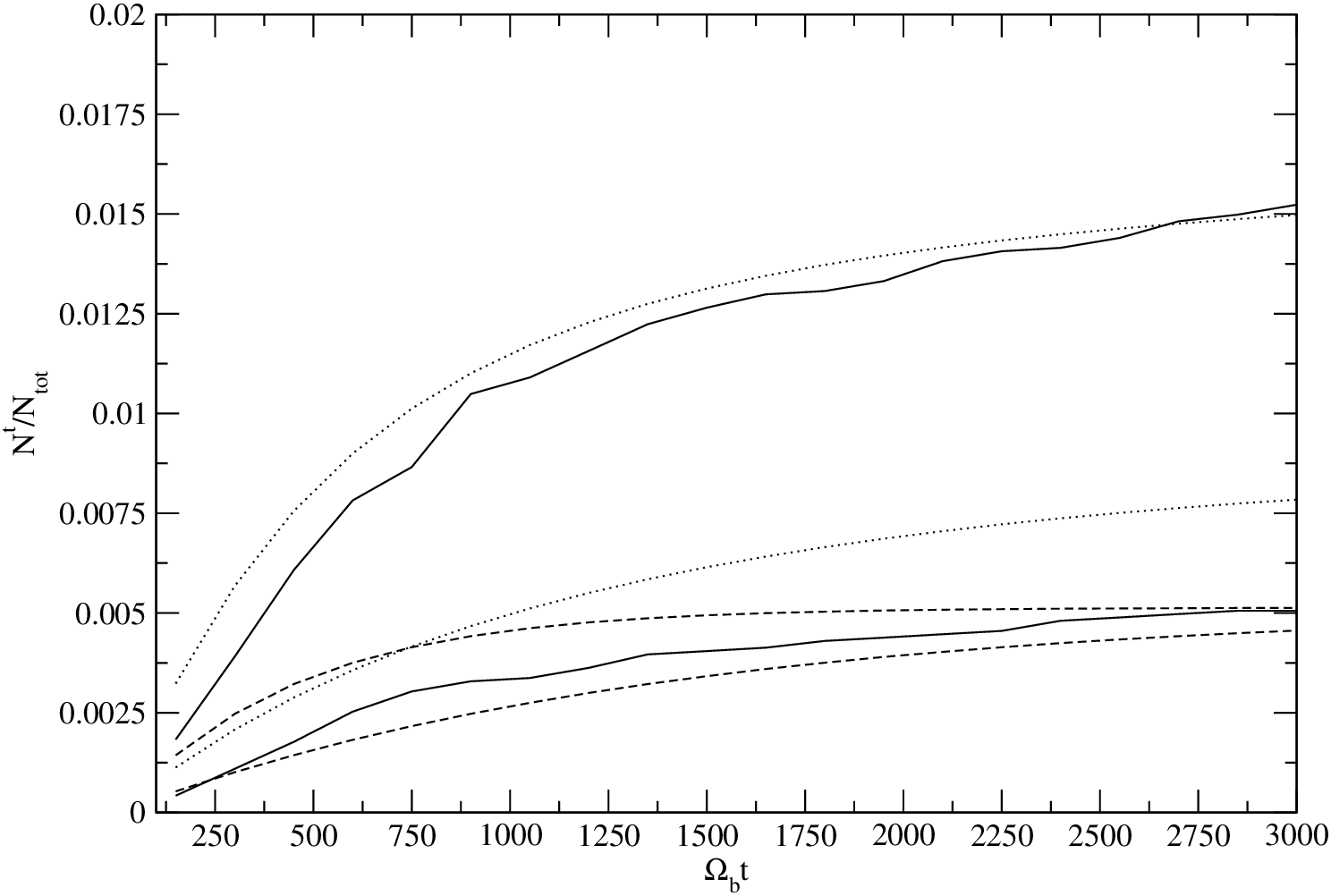}
\caption{Same as Fig.~4 and Fig.~5 but calculated for $q=1\cdot 10^{-2}$.}
\end{figure}
     
In Figs.~8, 9 and 10 we show the dependencies of numerical and
theoretical values of the probability for the particle to be supplied
to the loss cone calculated for realistic values of the loss cone
$l_{T}=1\cdot 10^{-3}$ and $l_{T}=3\cdot 10^{-3}$. The mass ratio is
equal to $0.1$, $3\cdot 10^{-2}$, $1\cdot 10^{-2}$ in Fig.~8, Fig.~9
and Fig 10, respectively. The general behaviour of the curves is
similar in all cases. The curves corresponding to $l_{T}=3\cdot
10^{-2}$ are better described by the theoretical curve representing
the total probability $\prob_{\rm tot}(t)$. On the other hand the
curves corresponding to a very small $l_{T}=1\cdot 10^{-2}$ are much
better described by the curve representing the probability
$\prob_{1}(t)$. Note, that the agreement between the theoretical and
numerical values is very good in the case $q=0.1$, and the theoretical
and numerical curves almost coincide in this case. The disagreement
grows with decrease of the mass ratio $q$, and in the case $q=1\cdot
10^{-2}$ is about $\sim 10$ per cent.  Thus, in the case of a very low
size of the loss cone the probability for the particle to be supplied
to the loss cone is mainly determined by the effect of oscillations of
the angular momentum on the binary orbital time scale, and
respectively, by equation (49). In fact, the numerical results
demonstrate that for a rough estimate in the case of a large ratio
$l_{T}/q$ we can use the expression for the probability determined by
the Kozai effect (equation 41) while in the opposite case of the small
ratio $l_{T}/q$ the expression (49) determined by the effect of the
oscillations of the angular momentum. As we have discussed above, this
conclusion is well motivated from theoretical point of view.
   
\section{Astrophysical applications}

\subsection{The size of the tidal loss cone}

We do not concentrate in this paper on complicated issues related to
the problem of tidal disruption of a star, and use the simplest
criterion of the tidal disruption event assuming that all stars having
pericentres $R_{p}$ smaller than the so-caller `tidal' radius $R_{T}$
are tidally disrupted:
$$R_{p} < R_{T}=\left({M\over m_{st}}\right)^{1/3}R_{st}\approx
1.5\cdot 10^{13}M_{7}^{1/3}\cm, \eqno 71$$ where we assume that all
stars have solar masses and radii $R_{st}\approx 7\cdot 10^{10}\cm$ and
$M_{7}=M/10^{7}M_\odot$. A simple estimate (Hills, 1975, Beloborodov et al 1992) 
shows that
when the mass of the central black hole is larger than the `critical
mass' $M\sim 1\cdot 10^{8}M_\odot$ the stars are swallowed by the
black hole without disruption, and therefore the range of the black
hole masses interesting for the tidal problem is limited from above:
$M < 1\cdot 10^{8}M_\odot$.\footnote{Note, however that a rotating
black hole of mass $\sim 10^{9}M_\odot$ can actually disrupt the stars
due to effects of General Relativity (Beloborodov et al 1992, Ivanov,
Chernyakova \& Novikov, 2003). The tidal disruption rate is suppressed
in this case by a geometrical factor, and we neglect this effect later
on.}  On the other hand, at the present time there is no observational
evidences of black holes masses within the intermediate mass range $M
< 10^{5} M_\odot$ (e.g., Kormendy \& Gebhardt 2001 and references
therein), and therefore we assume that our parameter $q$ cannot be
smaller than $\sim 10^{-3}$. The specific angular momentum $L_{T}$
associated with $R_{T}$ is obtained from the standard Keplerian
relation $L_{T}=\sqrt {2GMR_{T}}$. Following from the previous
Sections, in our calculations we use the dimensionless angular
momentum variable $l_{T}=L_{T}/\sqrt{GMD}$, where $D$ is the
separation distance between the two black holes.  The numerical value
of $l_{T}$ can easily be obtained from equation (71):
$$l_{T}\approx 3\cdot 10^{-3}M_{7}^{1/6}{(\tilde D r_{1})}^{-1/2}, \eqno 72$$
where $\tilde D=D/r_{a}$ and $r_{1}=r_{a}/1\pc$.  We see a rather weak
dependence of the numerical value of $l_{T}$ on the parameters,
especially on the mass of the central black hole.

\subsection{The dynamical friction}

When the secondary black hole enters the cusp it interacts with the
stars in the cusp and continues to spiral down to the centre of mass
due to the effect of dynamical friction. As a result the separation
distance between the primary and the secondary black holes is a
decreasing function of time $D(t)$. To estimate the characteristic
time scale associated with this process we have to specify the
distribution function of the stars in the cusp as well as the orbital
parameters of the binary. As we have mentioned above we assume that
the binary is circular and use the isotropic distribution function (1)
with $p=0$. In this case the evolution law of the binary separation
distance is especially simple:
$$D(t)=r_{a}\exp{(-t/T_{d.f.})}, \eqno 73$$
where $r_{a}$ is the radius of the cusp (see Section 2), and the characteristic evolution time scale $T_{d.f.}$
can be easily obtained from the expressions given in Chandrasekhar 1960:
$$T_{d.f.}=
{1\over 3\pi}q^{-1}\Lambda_{c}^{-1}P_{a}\approx 2\cdot 10^{4}
\left({1\cdot 10^{-2}\over q}\right)M_{7}^{-1/2}r_{1}^{3/2}\yr, \eqno 74 $$
where $q=m/M$ is the binary mass ratio, $\Lambda_{c}\sim 15$ is the
so-called Coulomb logarithmic factor, $r_{1}=r_{a}/1\pc$, and $P_{a}$
is a typical period of the binary with separation distance $D\sim
r_{a}$
$$P_{a}=2\pi \sqrt {{r_{a}^{3}\over GM}}\approx
3\cdot 10^{4}M_{7}^{-1/2}r_{1}^{3/2}\yr. \eqno 75$$
The evolution law (73) tells that the characteristic evolution time
scale $D/\dot D$ is independent of the distance $D$ and is always
approximately equal to $T_{d.f}$.

Equation (73) has its limits of applicability. When the binary evolves
to the radius $r_{m}=r_{a}q^{2/3}$ where the total mass of the stars
inside the binary orbit is of order of the secondary mass $m$ the
evolution is halted (Begelman, Blanford \& Rees 1980).\footnote{In
fact, the evolution may proceed due to the effect of gas dynamic drag
exerted on the binary by e.g., an accretion disc (Ivanov, Polnarev \&
Papaloizou 1999). We do not consider this possibility in the present
paper.} So we have $\tilde D > q^{2/3}$.  On the other hand, the
expression (73) is approximately valid only when the time $T_{d.f.}$
is larger than the binary orbital period
$$P_{b}=2\pi \Omega_{b}^{-1}=P_{a}{\tilde D}^{3/2},$$
hence $\tilde D < {({1\over 3\pi q \Lambda_{c}})}^{2/3}$.  When the
mass ratio of the binary is larger than a certain value $q_{\rm crit}$
both conditions are incompatible with each other and our simple
description of the dynamical friction becomes qualitatively
invalid. The value $q_{\rm crit}$ can be found by equating $T_{d.f.}$
and $P_{b}$ at the radius $r_{m}$
$$q_{\rm crit}={1\over \sqrt{3\pi \Lambda_{c}}}\sim 0.1, \eqno 76$$
where we recall that $M_{st}=M\tilde D^{3/2}$ (equations (4) and (5)
with $p=0$) and use $\Lambda_{c}=15$.  When the mass ratio $q$ is in
the range: ${1\over 3\pi \Lambda_{c}} \sim 1\cdot 10^{-2} < q < q_{\rm
crit}\sim 0.1$ equation (73) is approximately valid only for the
dimensionless separation distance $\tilde D$ smaller than
$$\tilde D_{\rm crit}={1\over {(3\pi \Lambda_{c} q)}^{2/3}}=
{q_{\rm crit}^{4/3}\over q^{2/3}}
\approx 0.2 \left({q_{\rm crit}\over q}\right)^{2/3}. \eqno 77$$

\subsection{The tidal disruption rate}

The expression for the tidal disruption rate, $\dot M_{t}$ is
determined by the ratio of the characteristic time scale of the
secular evolution given by equations (68) and (70) to the time scale
of the dynamical friction $T_{d.f}$. When this ratio is small, for a
given value of the separation distance $D$ almost all stars with
semi-major axes of the order of $D$ and sufficiently small values of
the $z$ component of angular momentum have a chance to be tidally
disrupted.  In this case $\dot M_{t}$ can be expressed in terms of the
time scale $T_{d.f}$, the probabilities $\prob_{\rm tot}$ and
$\prob_{tot,1}$ given by equations (67) and (69), and the total mass
of the stars within the sphere of radius $D$, $M_{st}(D)$. When $l_{T}
\gg q$ we have
$$\dot M_{t}(D) \approx {\prob_{\rm tot} M_{st}(D)\over T_{d.f.}}, \eqno 78$$
and in the opposite case $l_{T} \ll q$ we should use $\prob_{tot,1}$
instead of $\prob_{\rm tot}$ in equation (78). When the time scales of
the secular evolution are larger than $T_{d.f.}$ the bulk of the stars
have no time to be tidally disrupted, and the tidal disruption rate is
determined by equation (66):
$$\dot M_{t}(D)\approx \dot \prob M_{st}(D). \eqno 79$$

As follows from equation (72) the size of the loss cone increases when
the separation distance $D$ decreases.  That means that, in principal,
the ratio $R=l_{T}/q$ is a function of scale. However, it is easy to
see from equation (72) that when $q > 1\cdot 10^{-2}$ we have $R < 1$
for the all scales of interest $r_{m} < D < r_{a}$, and when $q <
3\cdot 10^{-3}$ we have $R > 1$ for the relevant scales. For
simplicity we are going to consider separately two possible cases: 1)
the case of a large mass ratio $q > 1\cdot 10^{-2}$ and 2) the case of
a small mass ratio $q < 3\cdot 10^{-3}$.

In the case of the large mass ratio it can be easily shown that the
time scale $T_{d.f}$ is always smaller than or of the order of the
time scale $T_{*,1}$ given by equation (70). Accordingly, we can use
equation (79) and obtain the explicit expression for $\dot M_{t}$ with
help of equations (66), (72) and (75):
$$\dot M_{t}\approx 4\cdot 10^{-1}
\left({q\over 10^{-2}}\right)^{2/3}
\mu^{-1/3}M_{7}^{5/3}r_{1}^{-2}{M_\odot \over \yr},\eqno 80$$
where we use the radial variable $\mu =M_{st}(D)/m$ instead of $D$. It
is obvious that the variable $\mu$ takes its values in the range $1 <
\mu < q^{-1}$. $\dot M_{t}$ decreases with increase of $\mu$ and
therefore the innermost part of the cusp $\mu=1$ gives the highest
tidal disruption rate. Note, however, that this dependence is rather
slow $M_{t} \propto \mu^{-1/3}$.

In the case of the small mass ratio there is a scale $\mu_{*}$ where
the secular time $T_{*}$ given by equation (68) is equal to $T_{d.f}$:
$$\mu_{*}\approx \left({3\cdot 10^{-3}\over q}\right)^{2}. \eqno 81$$
The tidal disruption rate is given by equation (80) for $\mu >
\mu_{*}$. In the opposite case we use equation (78) with the result
$$\dot M_{t}\approx 2\cdot 
10^{-1}\left({q\over 3\cdot 10^{-3}}\right)^{5/3}\mu^{1/6}M_{7}^{5/3}r_{1}^{-2}
{M_\odot\over \yr}. \eqno 82$$  
The tidal disruption rate is maximal at $\mu=\mu_{*}$:
$$\dot M_{\rm max}\approx 2\cdot 
10^{-1}\left({q\over 3\cdot 10^{-3}}\right)^{4/3}M_{7}^{5/3}r_{1}^{-2}
{M_\odot\over \yr}. \eqno 83$$  

Estimates of the tidal disruption rate in the astrophysical systems
containing a single supermassive black hole give the value of the rate
smaller than or of the order of $\dot M_{*}\sim 10^{-4}M_\odot/\yr$
(e.g., Magorrian \& Tremaine 1999, Syer \& Ulmer 1999 and references
therein). Equating $\dot M_{*}$ to $\dot M_{\rm max}$ we can find the
minimal mass ratio, $q_{\rm min}$ sufficient for the tidal disruption rate
determined by the presence of the binary to surpass the standard value
$$q > q_{\rm min}\approx 10^{-5}r_{1}^{3/2}M_{7}^{-5/4}. \eqno 84$$

There is another important limitation on our formalism determined by the
relativistic effect of Einstein apsidal precession of the stellar 
orbits (Hopman $\&$ Alexander 2005). To estimate the importance of this 
effect we use the standard expression for the change of the argument of
pericentre due to this effect:
$$\delta \omega_{rel} ={6\pi GM\over c^{2} \epsilon_{T}^{2} a},$$
where $\epsilon_{T}=l_{T}/\sqrt {\tilde a}$ is the value of the 
parameter $\epsilon$ associated with the 'tidal' angular momentum 
$l_{T}$.
We calculate the corresponding precession rate 
$\dot \omega_{rel}=\delta \omega_{rel}/P_{orb}$, and compare it with
the precession rate determined by the Kozai effect in the limit of
high eccentricity: $\dot \omega_{K}\sim 5/(T_{K}\epsilon_{T})$ 
(see equation 12). 
We find that the Einstein precession is not important provided that
the mass ratio is sufficiently high:
$$q > q_{rel}\sim 1.5\cdot 10^{-3}{M_{7}^{5/6}\over \sqrt{\tilde D r_{1}}},
\eqno 85$$ 
where we set $a=D/2$ for a 'typical' stellar orbit.'
  
Finally, let us estimate the duration of the phase of the enhanced
tidal disruption rate due to the presence of the secondary. The
corresponding time scale, $T_{T}$ follows from equation (73)
$$T_{T}=\ln{(r_{a}/r_{m})}T_{d.f.}={2\over 3}\ln{(q^{-1})}T_{d.f}. \eqno 86$$
We also neglect the difference between $D_{\rm crit}$ (equation 77)
and $r_{a}$ which enters in the expression under the
logarithm. Substituting equation (74) in (86) we find that $T_{T}$ is
rather short:
$$T_{T}=6\cdot 10^{4}\left(1+{1\over 4.6}\ln{10^{-2}\over q}\right)
M_{7}^{-1/2}r_{1}^{3/2}\yr. \eqno 87$$

\section{Conclusions and Discussion}

We have shown in this paper that the tidal disruption rate in a
galactic cusp containing a circular black hole binary can be as large
as $\sim 0.1$--$1 M_\odot/\yr$ for a cusp size of the order of
$1\pc$. Cusps of this size have been observed in e.g. our own Galaxy and
 M32 (e.g. Kormendy $\&$ Gebhardt 2001 and references therein). 
In principle, the rate of gas 
supply to the centre of the
galaxy is sufficient to power a quasar or a powerful AGN. However, the
estimated duration of the stage of the enhanced tidal disruption rate
$\sim 10^{5} \yr$ seems to be much shorter than the expected life time
of the central engine in quasars. Also, note that the tidal disruption
rate sharply decrease with increase of the cusp size $r_{a}$: $\propto
r_{1}^{-2}$, where $r_{1}=r_{a}/1pc$. 
The obsevations suggest that the cusp size grows
with mass of the central black hole with the typical sizes in the range 
$10-30\pc$ for the masses of central black holes 
of the order of $10^{7}-10^{8}M_{\odot}$
(e. g. Kormendy $\&$ Gebhardt 2001).   
So for an extended cusp of size $\sim 10\pc$ we get
rather moderate values of the tidal disruption rate of the order of
$1\cdot 10^{-3}-1\cdot 10^{-2}M_\odot/\yr$.

The tidal disruption events in a system containing a supermassive
black binary black hole could have observational consequences on its
own. A single tidal disruption event could manifest itself as sharp
rise of luminosity of a galactic centre on time scale $\sim 1\yr$
(Gurzadian \& Ozernoi 1981, Lacy et al 1982, Rees 1988, Komossa et al
2003 and references therein).  The repeating luminosity flares in a
particular galaxy could suggest that there is a binary black hole in
the centre. On the other hand, one can expect the increased
probability of the tidal disruption events in the supermassive binary
black hole candidates (e.g., Sillanpaa et al 1988, Valtaoja et al
2000).

We assume that the stellar distribution has a pronounced density
increase toward the centre. This assumption is well motivated from
theoretical point of view and also has been confirmed by analysis of
the distribution of stars inside our own Galaxy (e.g., Alexander 1999)
and some other galaxies (e.g., Faber et al.\ 1997).  However, after the
intervention of the secondary black hole and coalescence with the
primary, a large fraction of the stars is expelled from the centre,
and the density increase is smeared out. This may explain rather shallow
density profiles observed in the centers of bright galaxies 
(e.g., Faber et al.\ 1997) Thus, subsequent coalescence
events of two black holes within the centre of the same galaxy would
be accompanied by smaller numbers of tidally disrupted stars.

We have considered only the `inner problem' where all stars going to
the loss cone were assumed to be inside the binary orbit and showed
that for the inner problem the bulk of the stars going to loss cone
would have had a small value of the projection of the angular momentum
on the axis perpendicular to the binary orbital plane before
intervention of the secondary in the cusp. The mechanism responsible
for the supply of the stars into the loss cone is analogous to the
well known Kozai effect in celestial mechanics. However, the important
effect of the oscillation of the angular momentum on the time scale of
the order of the binary orbital period should also be taken into
account in order to obtain the correct expression for the tidal
disruption rate in a binary with a sufficiently large mass
ratio. Also, we take into account the apsidal precession determined by
gravitational potential of the stellar cluster itself and develop the
analytical theory of secular evolution in the combined gravitational
field of the binary and the stellar cluster. This theory might be
useful in other astrophysical applications, such as e.g., the theory
of dynamics of the triple stars (e.g., Kiseleva, Eggleton \& Mikkola
1998 and references therein), and the theory of dynamics of
extra-solar planets around the binary or a single star (e.g.,
Papaloizou \& Terquem 2001).

The tidal disruption rate may be 
significantly enhanced if we take into account `the outer' stars with semi-major axes
larger than the half of the binary separation distance. 
The evolution of these stars in the binary gravitational field
is mainly governed by two processes: 
the secular non-resonant interaction with the binary similar to the interaction
discussed in our paper, and the 
effect of close encounters with the secondary which change the orbital elements
of the star in a complicated chaotic way. 
Unfortunately, this problem looks very complicated for a thorough theoretical
treatment and numerical calculations are very important in this case.      
 
Our estimate of the rate of supply of the stars to the tidal loss cone
relied upon the effect of the secular increase of the eccentricity. We
have analysed this effect in an idealised situation where only the
Keplerian gravitational field of the binary and corrections due to the
stellar cluster have been taken into account and the binary has had a
circular orbit with the fixed separation distance between the primary
and the secondary. In a real astrophysical situation the binary could
have a substantial eccentricity (e.g., Quinlan $\&$ Hernquist 1997,
Aarseth 2003) and the orbital
parameters of the binary are evolving with time due to the effect of
dynamical friction. Therefore, a very important question arises: to
what extent is the effect of the secular increase of the eccentricity
of the stellar orbits sensitive to possible corrections due to
non-zero eccentricity and to a change of the orbital elements of the
binary?  First we would like to point out that when the eccentricity
of the binary is not very large the effects determined by the
eccentricity seem to be rather insignificant for the inner problem
considered above. Indeed, in the quadrupole approximation used in our
analysis the gravitational field of the eccentric binary is formally
equivalent to the gravitational field of the circular binary provided
that we use the parameter of the binary orbit
$p_{B}=a_{B}(1-e_{B}^{2})$ instead of the separation distance $D$ in
equation (14) defining the characteristic time-scale of the Kozai
effect, $T_{K}$ (e.g Innanen, Zheng, Mikkola and Valtonen 1997).  Of
course, in the case of a highly eccentric binary the influence of the
next order corrections in the small parameter $a/D$ would be more
prominent, but we expect that the qualitative character of the secular
evolution would remain unchanged until the condition $a/D < 1$ holds
for the stellar orbits of interest. When the quadrupole approximation
is valid, a slow change of the orbital parameters of the binary also
seems to be unimportant. In this case this time dependence enters the
equations of the secular evolution (10--13) only through the time
dependence of the quantity $T_{K}(t)$ in front of the time
derivatives. Introducing the new time variable $t_{1}=\int dt/T_{K}$
we can eliminate this time dependence, and the secular evolution is
the same in terms of the new time variable $t_{1}$ as it was in the
case of the non-evolving binary orbital elements.

In this Paper we consider the black holes as point like sources of 
Newtonian gravity. However, as we discussed in the previous Section,
the relativistic effect of Einstein apsidal precession (and, possibly,
the effect of gravitomagnetic precession) could be important. A study
of these effect could be a very interesting problem deserving a separate
study.

In conclusion, we would like to point out that for a quantitative
calculation of the tidal disruption rate our approach is perhaps too
oversimplified, and a thorough numerical study is needed.  However,
our model can be used for the order of magnitude estimates discussed
above. It can also be used for guidance on the choice of parameters in
numerical studies.

\section*{Acknowledgements}
It is a pleasure to thank J.C.B. Papaloizou for many important and
fruitful discussions and comments. We also thank Clovis Hopman for useful remarks. 
The authors acknowledge support from
PPARC through research grant PPA/G/02001/00486.

\section*{Appendix A: The precession rate}

Here we derive expressions for the apsidal precession rate of a stellar orbit 
induced by the gravitational field of the stellar cusp.
As follows from the discussion above the mass of the stars in the cusp $M_{st}$ and accordingly the gravitational potential 
$\Phi_{st}$ depend on the radius $r$ as a power of
the radius
$$M_{st}=M\left({r\over r_{a}}\right)^{3/2-p}, \quad
\Phi_{st}={GM\over (1/2-p)r_{a}}\left({r\over r_{a}}\right)^{1/2-p}, \eqno A1$$ 
Assuming that $r \ll r_{a}$ we can consider $\Phi_{st}$ as a perturbation. In this case the change $\delta \omega $
of the angle $\omega$ per one orbital period is given
by the expression (e.g., Landau, Lifshitz 1969)
$$\delta \omega=2{d\over dL}
\left({1\over L} \int^{\pi}_{0}d\phi (r^{2}\Phi_{st})\right), \eqno A2$$
where the radius $r$ is determined by the standard Keplerian relation
$$r={L^{2}\over GM(1+e\cos (\phi))}. \eqno A3$$ 
Substituting (A1) in (A2) and introducing the new variable $\epsilon=\sqrt{(1-e^{2})}$ instead of the eccentricity $e$ we obtain
$$\delta \omega = {2\over s-1}\left({a\over r_{a}}\right)^{s}{d\over d\epsilon}(\epsilon^{2s+1}J), \eqno A3$$
where $s=3/2-p$, $a$ is the semi-major axis, we use $L=\sqrt{(GMa)}\epsilon$, and
$$J=\int^{\pi}_{0}{d\phi \over \left(1+e\cos \phi\right)^{s+1}}, \eqno A4$$

When $s \ne n/2$ where $n$ is an integer, the integral $J$ can be 
expressed in terms of the Legendre function $P_{s}(x)$ as
$$J=\pi \epsilon^{-(s+1)}P_{s}(\epsilon^{-1}), \eqno A5$$
and $P_{s}(x)$ is expressed itself in terms of the hyper-geometric function $F(\alpha; \beta; \gamma; x)$ as
$$P_{s}(x)={2^{s}\sqrt \pi}{\Gamma(s+1/2)\over \Gamma(s+1)}x^{s}
F((1-s)/2;-s/2;1/2-s; 1/x^{2})+$$  
$${\tan(s \pi)\over 2^{s+1}\sqrt \pi}{\Gamma(s+1)\over \Gamma(s+3/2)}x^{-s-1}
F(s/2+1;(s+1)/2;s+3/2; 1/x^{2}), \eqno A6$$
where $\Gamma(x)$ is the gamma function.

We are interested in $s=3/2$ corresponding to $p=0$. In this case the integral $J$ can be expressed in terms of
the complete elliptic integrals as
$$ J={2\over 3}{\sqrt {(1+e)}\over \epsilon^{4}}
\left(4{\bf E}\left(\sqrt{{2e\over 1+e}}\right)
-(1-e){\bf K}\left(\sqrt{2e\over 1+e}\right)\right). \eqno A7$$

When eccentricity is small, the expressions are simplified significantly. In the leading order in the small parameter
$\epsilon$ we obtain from equations (A3), (A5) and (A6)
$$\delta \omega=-K\left({a\over r_{a}}\right)^{s}\epsilon, \eqno A8$$
where
$$K=\sqrt \pi {2^{s+1} \over (2s-1)}{\Gamma(s+1/2)\over \Gamma(s)}. \eqno A9$$
Substituting $p=3/2-s$ in equation (A9) we obtain
$$K= 2^{3/2-p}\sqrt \pi {\Gamma(1-p)\over \Gamma(3/2-p)}. \eqno A10$$
In the case $p=0$ ($s=3/2$) we use equations (A3) and (A7) to obtain
$$\delta \omega=-4\sqrt 2 \left({a\over r_{a}}\right)^{3/2}\epsilon. \eqno A11$$
Note that equation (A11) follows from equations (A8) and (A9) when $s=3/2$.

\section*{Appendix B: Evolution of the angular momentum}

In the quadrupole approximation the rate of change of square of the angular momentum of a star on an elliptic 
orbit due to gravitational perturbation of a perturber of mass $m$ is given by the expression:
$${d\over dt}L^{2}={6Gm\over D^{5}}
\lbrace ({\vec r}\cdot {\vec r_{p}})(r^{2}(\dot {\vec r} \cdot {\vec r_{p}})-({\vec r}\cdot {\vec r_{p}})
({\vec r}\cdot \dot {\vec r}))\rbrace, \eqno B1$$ 
where $D$ is the separation distance, $\vec r$ is the radius vector of the star with coordinates $(x(t),y(t),z(t))$
and $\vec r_{p}$ is the radius vector of the perturber with coordinates $(x_{p}(t)=D\cos (n),y_{p}=D\sin (n),0)$, where
$n=\Omega_{p}t+n_{0}$ and $\Omega_{p}=\sqrt{{GM\over D^{3}}}$. In the expression in the braces in (A1) we can use the
unperturbed values of $\vec r(t)$ and $\vec r_{p}(t)$. Since the orbit of the star is in general inclined with respect to
the orbital plane of the perturber, it is convenient to introduce new coordinates associated with the 
orbital plane of the star where the radius vector of the star has two non-zero components: 
$(\tilde x(t),\tilde y(t), 0)$, 
and relate these coordinates to the 
coordinates $(x(t),y(t),z(t))$ with help of the Euler angles.
We have 
$$x=(\cos (\omega )\cos (\chi ) - \sin (\omega) \sin (\chi) \cos (i))\tilde x
-(\sin (\omega )\cos (\chi)+\cos (\omega)\sin (\chi)\cos (i))\tilde y, \eqno B2$$
$$y=(\cos (\omega)\sin (\chi)+\sin (\omega)\cos (\chi)\cos (i))\tilde x-
(\sin (\omega)\sin (\chi)-\cos(\omega)\cos (\chi)\cos (i))\tilde y, \eqno B3$$
where $i$ is the inclination, $\omega $ is the argument of pericentre and $\chi $ is
the longitude of ascending node. $\tilde x(t)$ and $\tilde y(t)$ are given by the standard Keplerian 
expressions
$$\tilde x(t)= a(\cos (\xi)-e), \quad \tilde y(t)=a\epsilon \sin(\xi), \eqno B4$$
where $a$ is the semi-major axis, $e$ is the eccentricity, $\epsilon=\sqrt{(1-e^{2})}$ and $\xi$ is the 
eccentric anomaly. Substituting (B2-B4) into the expression in the braces in equation (B1) we have
$$A=-\epsilon D^{2}a^{4}B\dot \xi, \eqno B5$$
where the variable $A$ represents the expression in the braces and 

$$B=(1-e\cos (\xi))((\cos (\xi)-e)\epsilon \sin
(\xi)(\alpha^{2}-\beta^{2})+((\cos(\xi)-e)^{2}
-\epsilon^{2}\sin^{2} (\xi))\alpha \beta), \eqno B6$$

$$\alpha=\cos (\omega )\cos (\phi )+\sin (\omega )\sin (\phi)\cos (i), \quad
\beta=\sin (\omega) \cos (\phi) -\cos (\omega)\sin (\phi) \cos (i), \eqno B7$$
and $\phi =n -\chi$.

After averaging of (B5) over the orbital period of the star, $P_{orb}$, we obtain
$$\bar A = {1\over P_{orb}}\int^{2\pi}_{0}d\xi A=-{5\over 2} (\alpha \beta)\epsilon e^{2}D^{2}a^{4}\Omega, \eqno B8$$
where $\Omega=\sqrt{{GM\over a^{3}}}$. Substituting (B8) into
(B1) and taking into account the standard relation $L=\epsilon \sqrt{GMa}$, we obtain: 
$${d\over dt}L=-{15\over 2}(\alpha \beta)e^{2}({a \over D})^{3}{Gm\over a}, \eqno B9$$
where 
$$\alpha \beta=-{1\over 2}\sin (2\phi)\cos (i) \cos (2\omega)+
{1\over 4}\sin (2\omega)(\sin^{2} (i)+
\cos (2\phi)(1+\cos^{2} (i))). \eqno B10$$
After averaging over the binary period equation (B9) must agree with
equation (10). Taking into account that the averaged value of $\alpha
\beta $ is equal to ${1\over 4} \sin (2\omega) \sin^{2} (i)$ and using
$e=\sqrt{ 1-{L^{2}\over GMa}}$, we obtain:
$${d \over dt}e={15\over 8}q\epsilon e \sin (2\omega) \sin^{2} (i)
\left({a\over D}\right)^{3}\Omega , \eqno B11$$
where $q=m/M$. It is easy to see that equation (B11) is identical to equation (10). 

Now let us estimate the amplitude of oscillations of the angular
momentum $\Delta L$ on time scale comparable to the orbital period of
the perturber. Neglecting for a moment these oscillations and assuming
that the eccentricity of the star due to evolution of the orbital
elements on the `slow' Kozai time scale is close to its maximal value,
and accordingly, the angular momentum is close to its minimal value
$L_{\rm min}$, it follows from the discussion above that the
inclination $i_{\rm min}$ is minimal and $\omega \approx \pm {\pi
\over 2}$. Therefore, we have $\alpha \beta\approx -{1\over 2}\cos
(i_{\rm min}) \sin (2\phi)$ from $(A10)$, and equation $(B9)$ takes
the form:
$${d\over dt}L\approx -{15\over 4}
\cos (i_{min})\left(a\over D\right)^{3}
\left({Gm\over a}\right)\sin (2\phi). \eqno B12$$
Integrating (B12) we obtain:
$$L\approx L_{\rm min}+\Delta L \cos (2\phi), \eqno B13$$
where
$$\Delta L ={15\over 8}\cos (i_{\rm min})
q\left({a\over D}\right)^{2}\sqrt{GMD}. \eqno B14$$ 

\bsp

\label{lastpage}

\end{document}